\documentclass[superscriptaddress,showpacs,aps,prb,twocolumn]{revtex4}

\usepackage{times}
\usepackage{amsmath}
\usepackage{amssymb}
\usepackage{graphicx}
\usepackage{pst-all}

\begin{document}

\title{The non-Abelian Interferometer}

\author{Waheb~Bishara}
\affiliation{Department of Physics, California Institute of Technology, Pasadena, California 91125, USA}

\author{Parsa~Bonderson}
\affiliation{Microsoft Research, Station Q, Elings Hall, University of California, Santa Barbara, CA 93106, USA}

\author{Chetan~Nayak}
\affiliation{Microsoft Research, Station Q, Elings Hall, University of California, Santa Barbara, CA 93106, USA}
\affiliation{Department of Physics, University of California, Santa Barbara, CA 93106, USA}

\author{Kirill~Shtengel}
\affiliation{Department of Physics and Astronomy, University of California at Riverside, Riverside, CA 92507}
\affiliation{Institute for Quantum Information, California Institute of Technology, Pasadena, California 91125, USA}

\author{J.~K.~Slingerland}
\affiliation{Dublin Institute for Advanced Studies, School of Theoretical Physics, 10 Burlington Rd, Dublin, Ireland}
\affiliation{Department of Mathematical Physics, National University of Ireland, Maynooth, Ireland}


\begin{abstract}
We consider the tunneling current through a double point-contact Fabry-P\'{e}rot interferometer such as used in recent experimental studies of the fractional quantum Hall plateau at filling fraction $\nu=5/2$.
We compare the predictions of several different
models of the state of the electrons at this plateau: the
Moore-Read, anti-Pfaffian, SU$(2)_2$ NAF, $K=8$ strong pairing, and $(3,3,1)$ states.
All of these predict the existence of charge $e/4$ quasiparticles,
but the first three are non-Abelian while the last two are Abelian. We give explicit formulas for the scaling of charge $e/2$ and charge $e/4$ quasiparticle contributions to the current as a function of temperature, gate voltage and distance between the two point contacts for all three models. Based on these, we analyze several possible explanations of two phenomena reported for recent experiments by Willett {\it et~al.}, namely halving of the period of the observed resistance oscillations with rising temperature and alternation between the same two
observed periods at low temperatures as the area of the interference loop is varied with a side gate.
We conclude that the most likely explanation is that
the observed alternation is due to switching between
even and odd numbers of charge $e/4$ quasiparticles
enclosed within the loop as a function of side gate voltage,
which is a clear signature of the presence of non-Abelian anyons. However, there are important features of the data which do not have a simple explanation within this picture.
We suggest further experiments which could help rule out
some possible scenarios. We make the corresponding predictions for future tunneling and interference experiments at the other observed second Landau level fractional quantum Hall states.
\end{abstract}

\date{\today}

\pacs{
71.10.Pm, 
73.43.-f, 
73.43.Jn  
05.30.Pr 
}
\maketitle

``With luck, we might see a non-abelian interferometer within a year.'' -- attributed to Kirill Shtengel, April 16, 2008 in \emph{Quantum computation: The dreamweaver's abacus}~\cite{Dreamweaver08}.

\section{Introduction}
\label{sec:introduction}

The observation~\cite{Willett87,Pan99} of a fractional quantum Hall (FQH) state at $\nu=5/2$ and suggestion~\cite{Greiter92} that the Moore-Read Pfaffian (MR) state~\cite{Moore91,Nayak96c,Read96}
might occur at this filling fraction
gave the first real indication that non-Abelian topological
phases of matter might actually occur in Nature. The striking feature of such new phases is that they possess quasiparticle excitations with exotic non-Abelian braiding statistics~\cite{Leinaas77,Goldin85,Fredenhagen89,Imbo89,Froehlich90,Imbo90,Bais92}. This property makes non-Abelian topological phases appealing for their potential use as intrinsically fault-tolerant media for quantum information processing~\cite{Kitaev03,Preskill98,Freedman98,Freedman02a,Freedman02b,Freedman03b,Preskill-lectures,Nayak08}.

Recent experimental studies of transport through a point contact in FQH systems at $\nu=5/2$ gave evidence that
there are charge $e/4$ quasiparticles in this state~\cite{Dolev08}
and found that the dependence of the current on voltage and temperature is most consistent~\cite{Radu08} with two particular
non-Abelian models: the anti-Pfaffian ($\overline{\text{Pf}}$) state~\cite{Lee07,Levin07} and the SU$(2)_2$ NAF (non-Abelian FQH) state~\cite{Wen91a,Blok92}.
However, these results are not conclusive because
the $(3,3,1)$ state~\cite{Halperin83}, which is Abelian,
also supports charge $e/4$ quasiparticles. It is
also roughly consistent with the voltage and temperature dependence of tunneling found in Ref.~\onlinecite{Radu08} and, in any case, one might expect non-universal physics to have a significant effect on the observed dependence.
Thus, there is a glaring need for experiments which directly probe the braiding statistics of quasiparticles.

In order to probe braiding statistics in FQH systems, one can use a double point-contact interferometer, as
proposed in Ref.~\onlinecite{Chamon97} for Abelian states and later considered for the $\nu=5/2$ state in Refs.~\onlinecite{Fradkin98,DasSarma05,Stern06a,Bonderson06a}. Such interferometers can play a crucial role in properly identifying which phase a FQH state is in by providing information about the topological $S$-matrix, a mathematical quantity related to the braiding statistics that is strongly characteristic of the topological order (for more details, see Ref.~\onlinecite{Bonderson06b}). Interferometers are also important for the implementation of topological quantum computation~\cite{Kitaev03,Freedman98} because they can be used for
the topological charge measurements necessary for readout of qubits~\cite{DasSarma05} and, through adroit manipulation, can even be used to implement computational gates~\cite{Bonderson08a,Bonderson08b}. Fortunately, there have been recent advances in realizing quantum Hall interferometers at integer filling~\cite{Ji03,Zhang09} and fractional filling in the lowest Landau level~\cite{Camino05a,Camino07a}. Even more recently, double point-contact interferometers have been experimentally implemented for the $\nu=5/2$ FQH state~\cite{Godfrey07,Willett08,Willett09a,Willett09u}.

In this paper, we study the signatures of non-Abelian statistics
which can be seen in a double point-contact interferometer
and discuss other effects which can mimic these signatures.
We propose further experiments which can help disentangle
the effects of non-Abelian statistics from Coulomb blockade
and disorder physics. The paper is structured as follows: In Section~\ref{sec:experiment}, we describe the basic features
of the experiment of Willett~{\it et al.}~\cite{Willett08,Willett09a,Willett09u}. In Section~\ref{sec:interpretations}, we explain three
different interpretations of this experiment: (a) non-Abelian interferometry, (b) Coulomb blockade,
and (c) possible explanations loosely grouped together because they depend on non-linear dependence
of the interferometer area on the side gate voltage $V_s$. In Section~\ref{sec:critical}, we criticize each of these three
interpretations. We argue that, while all three interpretations have problems, the problems with (b) and (c)
are more serious and these explanations are less likely to be correct. In Section~\ref{sec:proposals}, we propose
further experiments which might further strengthen or rule out interpretation (a). In Section~\ref{sec:discussion},
we comment on the implications of this experiment for topological quantum computation, assuming that
the non-Abelian interference interpretation is correct. In the two appendices, we give predictions for
interference experiments at other suspected non-Abelian fractions and we argue that the bare
backscattering amplitude for $e/4$ quasiparticles should be much larger than that for $e/2$ quasiparticles.

\section{The Experiment}
\label{sec:experiment}

\begin{figure}[b!]
\begin{center}
  \includegraphics[scale=1]{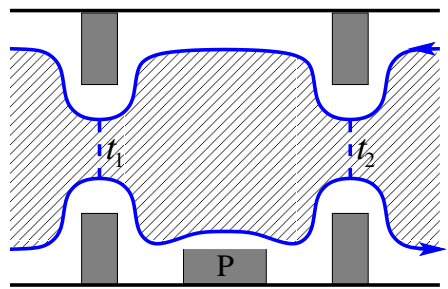}
  \caption{A double point-contact interferometer. Edge quasiparticles tunnel at two point-contacts with amplitudes $t_1$ and $t_2$, respectively. The interferometry area is changed by applying a voltage $V_{s}$ to a plunger gate $P$ that depletes the 2DEG beneath it. Quantum interference between the two paths manifests an observable signature of the Aharonov-Bohm effect and the braiding statistics (of the edge quasiparticle with the bulk quasiparticles in the central interferometry region) in the oscillation patterns of the tunneling current when the area is changed.}
  \label{fig:interf}
\end{center}
\end{figure}

In recent experiments, Willett~{\it et~al.}~\cite{Willett08,Willett09a,Willett09u}
measured the current through a double point-contact
device, depicted schematically in Fig.~\ref{fig:interf}.
As a function of magnetic
field $B$, the longitudinal resistance $R_L$
of the device has prominent minima at roughly
the $B$ values at which the $\nu=2, 7/3, 5/3$
and $5/2$ quantum Hall states occur in the bulk
(near, but not at, the point contacts).
At the minima corresponding to $\nu=5/2$ and $7/3$, the
longitudinal resistance is $R_{L} \simeq 200-300~\Omega$,
while at $\nu=2$ and $5/3$ it is $R_{L} < 50~\Omega$.
There are small oscillations with $B$ on top of these
large features, but these were not the focus of the experiment
since changing the magnetic field can change both the flux enclosed
and, possibly, the quasiparticle number, thereby making
it difficult to isolate the effect of braiding statistics. Instead,
a side gate voltage is varied, as shown in Fig.~\ref{fig:interf}.
As the side gate voltage $V_s$ is varied, $R_L$ oscillates with an
amplitude of roughly $2~\Omega$.

The period of the oscillations, $\Delta{V_s}$, is larger
at $\nu=5/3$ and $7/3$ than at $\nu=2$. This was interpreted
in the following way: it was assumed that the
principle effect of varying the side gate voltage is to change the
area of the interference loop between the two point
contacts and that they are related linearly by $\Delta A = c \Delta{V_s}$, where $c$ is essentially constant, even between different filling fractions. Thus, the oscillations are hypothesized
to be due to the Aharonov-Bohm (AB) effect, which implies
a period $\Delta A = ({e/e^*}){\Phi_0}/B$, where $e^*$ is the charge of the tunneling quasiparticle and $-e$ is the electron charge, and $\Phi_{0} = hc/e$ is the magnetic flux quantum.
Willett~{\it et~al.}~\cite{Willett09a,Willett09u} analyze their data to
find that the period at $\nu=5/3$ and $7/3$, normalized
by the corresponding magnetic fields, is three times larger
than at $\nu=2$: $(\Delta A)_{5/3}B_{5/3}\approx
(\Delta A)_{7/3}B_{7/3}\approx 3\cdot (\Delta A)_{2}B_{2}$.
Thus, they interpret their
findings as evidence that ${e^*}/e = 1/3$
at $\nu=5/3, 7/3$, assuming that the oscillation period
at $\nu=2$ reflects interference of ordinary electrons.
At $\nu=5/2$, two types of behavior are seen
at $25$~mK.
In some regions, which we will call type I,
$(\Delta A)^{\text{I}}_{5/2}B_{5/2}\approx
4\cdot (\Delta A)_{2}B_{2}$. In the regions of type II, $(\Delta A)^{\text{II}}_{5/2}B_{5/2}\approx 2\cdot (\Delta A)_{2}B_{2}$.
At $150$~mK, only one behavior is seen:
$(\Delta A)^{\text{II}}_{5/2}B_{5/2}\approx 2\cdot (\Delta A)_{2}B_{2}$. The type of oscillations observed for a region of $V_s$ were found to be reproducible throughout multiple scans over the period of $7$
days~\cite{Willett09u}. The type I oscillations in a given region sometimes exhibited a roughly $\pi$ phase shift from one scan to another. In the next section, we discuss several possible explanations
for the occurrence of these two periods at $25$~mK
and the disappearance of one of them at higher
temperatures at $\nu=5/2$.

\section{Interpretations}
\label{sec:interpretations}

\subsection{Non-Abelian Interference}

At first glance, these experimental results appear to be dramatically consistent with the predicted behavior of the proposed non-Abelian $\nu =5/2$ FQH states, particularly with that of the MR, $\overline{\text{Pf}}$, and SU$(2)_2$ NAF states, all of which have a non-Abelian fundamental quasihole with charge $e/4$. The basic assumption is that
as one changes the area of the interferometry region, one also occasionally
changes the number $n_q$ of charge $e/4$ quasiholes contained in the bulk within the interference loop. (For the purposes of this counting, charge $ne/4$ excitations, where $n \in \mathbb{Z}$, count as $n$ fundamental quasiholes.) Thus, changing the area will cause the edge current to exhibit interference behavior due to the AB effect,
modulated by occasional changes in the number of quasiparticles in the loop
and their concomitant braiding statistics. The interference term $I_{12}$ of the backscattered current due to lowest order tunneling of $e/4$ edge quasiholes is predicted to be~\cite{Stern06a,Bonderson06a}
\begin{equation}
\label{eq:nA_int_5_2}
I^{\left(e/4 \right)}_{12} \propto \left\{
\begin{array}{ll}
\cos \left( 2\pi \frac{\Phi}{4 \Phi_0} \mp \frac{n_{q} \pi}{4} +n_{\psi} \pi \right) & \text{for $n_q$ even} \\
 0 & \text{for $n_q$ odd}
\end{array}
\right.
,
\end{equation}
where the $-$ corresponds to the MR and SU$(2)_2$ states and the $+$ to the $\overline{\text{Pf}}$ state; and $n_{\psi} =0$ or $1$, depending on whether the contained quasiparticles are in a collective state corresponding to the $I$ or $\psi$ fusion channel. This interference exhibits the usual AB oscillations with period $\Delta A = 4 \Phi_0 /B$ corresponding to $e^*/e =1/4$, but also a striking complete suppression of this term that results from the non-Abelian braiding statistics of the edge quasiparticle with the bulk quasiparticles when $n_q$ is odd. Thus, as the area of the interferometry region is changed, and bulk quasiparticles enter or exit the interference loop, the non-Abelian states should see $\Delta A = 4 \Phi_0 /B$ oscillations switch on and off, as they do in going from the type I regions to the type II regions in the experiments
of Refs.~\onlinecite{Willett09a,Willett09u}.

The observed reproducibility of oscillation type regions in multiple scans~\cite{Willett09a,Willett09u} suggests that the bulk $e/4$ quasiparticles are pinned and do not move on the time scale of the experiment.
The observation of the oscillations in a given type I region being shifted by $\pi$ from one scan to the next also agrees with the expected behavior of non-Abelian states. Specifically, the collective state of several quasiparticles, some of which are inside and
some outside the interferometry loop, is decohered by the current of edge quasiparticles around the loop~\cite{Bonderson07a}. Hence, depending on the bulk quasiparticles entering or exiting the interferometry loop, the collective state of quasiparticles inside the interferometer may be randomized between $n_{\psi}=0$ and $1$ when $n_q$ is changed to an even value. (This is the same randomization that gives rise to a non-Abelian signature in the switching noise~\cite{Grosfeld06b}.)

There are two sources that could potentially contribute to $\Delta A = 2 \Phi_0 /B$ oscillations in the non-Abelian $\nu =5/2$ states. The first is tunneling of the Abelian $e/2$ edge quasiparticles~\cite{Bonderson07b,Bonderson07c}, which to lowest order gives the interference current~\cite{Chamon97}
\begin{equation}
\label{eq:I12e2}
I^{\left(e/2 \right)}_{12} \propto
\cos \left( 2\pi \frac{\Phi}{2 \Phi_0} - \frac{n_{q}\pi}{2} \right)
.
\end{equation}
The second possibility comes from higher-order tunneling processes where the interference path encircles the interferometry area twice. The resulting double pass interference term in the current coming from $2$nd order tunneling of $e/4$ edge quasiparticles is~\cite{Bonderson07b,Bonderson07c}
\begin{equation}
\label{eq:I1212e4}
I^{\left(e/4 \right)}_{1212} \propto \left\{
\begin{array}{ll}
 \cos \left( 2\pi \frac{\Phi}{2 \Phi_0} - \frac{n_{q}\pi}{2} \right) & \text{for $n_q$ even} \\
 \cos \left( 2\pi \frac{\Phi}{2 \Phi_0} - \frac{n_{q}\pi}{2} \pm \frac{\pi}{2} \right) & \text{for $n_q$ odd}
\end{array}
\right.
,
\end{equation}
where the $+$ corresponds to the MR state and the $-$ to the $\overline{\text{Pf}}$ and SU$(2)_2$ states. Of course, this $2$nd order contribution to the tunneling current will typically have much smaller amplitude, since it both incurs an additional tunneling probability factor and doubles the distance over which coherence must be maintained. For the interferometer of Refs.~\onlinecite{Willett08,Willett09a,Willett09u}, the quasiparticle tunneling probability at each point contact is approximately $P_1 \simeq P_2 \simeq .05$. This estimate is based on the relation~\cite{Fendley07a}
\begin{equation}
R_{xx} = \frac{h}{e^2} \frac{2}{5} \frac{P}{5-P}
\end{equation}
for point-contact tunneling of the half-filling edge modes at $\nu=5/2$, where $P \simeq P_1 + P_2$ here is roughly the sum of individual tunneling probabilities of the two point-contacts, and $R_{xx}\simeq 200$~$\Omega$ in Refs.~\onlinecite{Willett08,Willett09a,Willett09u}. Furthermore, there will generally be a suppression of the interference oscillation amplitudes that results from the loss of coherence, as well as from having unequal tunneling probabilities at the two point-contacts (which gives a suppression of $\frac{2 \sqrt{P_1 P_2} }{P_1 +P_2}$). Roughly speaking, we can define the suppression factor to be
\begin{equation}
Q \simeq \text{max} \left( I_{12} \right) / \left(I_1 +I_2 \right)
\end{equation}
as long as this is a small quantity. The observed oscillations in $R_{xx}$ have amplitude of approximately $2$~$\Omega$, indicating a suppression factor of $Q \simeq .01$. Higher order interference terms will be suppressed by higher powers of $P$ and $Q$, so combining these we find that the amplitude of double pass interference oscillations is expected to be roughly $.0005$ times that of the lowest order oscillation amplitude. Hence, the $\Delta A = 2 \Phi_0 /B$ oscillations, for which the amplitudes are of the same order of magnitude as that of the $\Delta A = 4 \Phi_0 /B$ oscillations, should be attributed almost entirely to the tunneling of $e/2$ edge quasiparticles. We emphasize that the $\Delta A = 2 \Phi_0 /B$ oscillations (from both sources) have an amplitude that is independent of $n_q$ (unlike the $\Delta A = 4 \Phi_0 /B$ oscillations), but pick up phase shifts when $n_q$ changes.

These two sources of $\Delta A = 2 \Phi_0 /B$ oscillations were not discussed in Refs.~\onlinecite{Stern06a,Bonderson06a} because it was assumed neither would have significant contributions to the tunneling current. For the double pass interference of $e/4$ quasiparticles, this appears to be a valid assumption, since higher-order tunneling processes are suppressed in the weak-backscattering regime. On the other hand, for interference of $e/2$ quasiparticles this assumption was based on such quasiparticles having less relevant tunneling operators than the $e/4$ quasiparticles. We will see in the following that there are several ways in which this line of reasoning can break down and permit the $e/2$ quasiparticles to have a contribution to the tunneling current oscillations that is comparable to that of the $e/4$ quasiparticles.

Combining these results, we see that tunneling of both non-Abelian $e/4$ quasiparticles and Abelian $e/2$ quasiparticles at the point contacts of the interferometer would produce a combined backscattered current with regions of type I, exhibiting a sum of both $\Delta A = 4 \Phi_0 /B$ and $\Delta A = 2 \Phi_0 /B$ oscillations, when $n_q$ is even, and regions of type II, exhibiting only $\Delta A = 2 \Phi_0 /B$ oscillations, when $n_q$ is odd. We also note that the bulk-edge coupling that occurs as a bulk $e/4$ quasiparticle approaches the edge gives the regions near transitions between type I and II oscillations the most potential for exhibiting non-linear and/or noisy behavior. The behavior of the interference current in the weak to strong coupling crossover as the quasiparticle approaches the edge was recently studied in Refs.~\onlinecite{Bishara09b,Rosenow09}, which found results not entirely inconsistent with the experimental data.

In order for interference to be observed, it is necessary that the
current-carrying excitations remain phase coherent. Even if
we neglect (irrelevant) interactions between the edge modes,
coupling to localized excitations in the bulk, and phonons,
there will still be thermal smearing of the interference pattern.
Consequently, as shown in Ref.~\onlinecite{Bishara08} (see also Ref.~\onlinecite{Fidkowski07c,Ardonne07a}), the amplitude of interference oscillation for double point-contact interferometers will be exponentially suppressed in
temperature and in the average length $L$ between point contacts along each edge
\begin{equation}
I^{\left(qp\right)}_{12}\propto e^{-T/T^{\ast}\!(L)}= e^{-L/L_{\phi}(T)} ,
\end{equation}
where the coherence length $L_{\phi}(T)$ and temperature ${T^*}(L)$
of edge excitations are given by
\begin{eqnarray}
\label{eq:coh_length}
L_{\phi}(T) &=& \frac{1}{2 \pi T }\left( \frac{g_c}{v_c} + \frac{g_n}{v_n} \right)^{-1} \\
\label{eq:coh_temp}
T^{*}(L) &=& \frac{1}{2 \pi L}\left( \frac{g_c}{v_c} + \frac{g_n}{v_n} \right)^{-1}
.
\end{eqnarray}
We can use these expressions, together with estimates of the charge and neutral edge mode velocities ($v_c \approx 5 \times 10^{4}$~m/s and $v_{n}  \approx 4 \times 10^{3}$~m/s) from numerical studies~\cite{Wan08a} of $\nu=5/2$ with pure Coulomb interactions on a disk~\footnote{The edge mode velocities are not universal quantities, and will generally be different for different samples, filling fractions, topological orders, etc. Lacking better physical estimates of the relevant velocities, we use these numerical velocity estimates to produce coherence length and temperature estimates for all the candidate $\nu=5/2$ states.} (the charged and neutral scaling exponents $g_c$ and $g_n$ are given in Table~\ref{Table:qps}), to estimate coherence lengths and temperatures for the charge $e/4$ and $e/2$ excitations in the various candidate states (the states are all the same, as far as the charge $e/2$ quasiparticle
is concerned). In Table~\ref{Table:coherences}, we give estimates of
coherence lengths at $T=25$~mK, and coherence temperatures for $L = 1~\mu$m. The temperature $T = 25$~mK is the lowest temperature at which the experiments of Refs.~\onlinecite{Willett08,Willett09a,Willett09u} were carried out, and $L = 1~\mu$m is the approximate interference path length on each side of the interferometry area determined in Refs.~\onlinecite{Willett08,Willett09a,Willett09u} to be $A \approx .2 \mu\text{m}^2$. We note that the observation of only type II oscillations at higher temperatures in Refs.~\onlinecite{Willett08,Willett09a,Willett09u} also excludes double pass interference of $e/4$ quasiparticles as the explanation for $\Delta A = 2 \Phi_0 /B$ oscillations, whereas it fits very nicely with the $e/2$ quasiparticle tunneling explanation.

\begin{table}
\[
\begin{array}{|l|c|c|c|c||c|}
\hline
e/4                         &  \text{MR}  &  \overline{\text{Pf}}/\text{SU$\left(2\right)_2$}  &  \text{K=8}  &  \text{(3,3,1)}  &  $e/2$    \\
\hline
L_{\phi} \text{ in $\mu$m}  &  1.4        &  0.5                                               &  19          &  0.7         &  4.8          \\
\hline
T^{*} \text{ in mK}         & 36          &  13                                                &  484         &  19          &  121          \\
\hline
\end{array}
\]
\caption{Estimated coherence lengths $L_{\phi}$ at $T = 25$~mK and coherence temperatures $T^*$ for $L = 1~\mu$m for the (relevant) $e/4$ quasiparticles of the candidate $\nu=5/2$ states, and the $e/2$ Laughlin-type quasiparticle for all these states. We use the velocity estimates $v_c \approx 5 \times 10^{4}$~m/s and $v_{n}  \approx 4 \times 10^{3}$~m/s from numerical studies~\cite{Wan08a}, while the temperature $T = 25$~mK and path length $L = 1~\mu$m are characteristic of the experiments of Willett {\it et al.}~\cite{Willett08,Willett09a,Willett09u}.}
\label{Table:coherences}
\end{table}

\subsection{Coulomb blockade}

If the region between the two point
contacts in the experiment of Willett {\it et al.}~\cite{Willett08,Willett09a,Willett09u}
was nearly an isolated puddle, the Coulomb
charging energy of the puddle would dominate the behavior
of the device. This might occur if the gates pinched off
the point contact too strongly.
Due to its isolation, the puddle must contain
an integer number of electrons. The electron number can
change when the gate voltage is increased
by enough to allow one additional electron into the puddle.
At this point, there is a peak in the longitudinal conductance
(which is also a peak in the longitudinal resistance,
since ${R_L}\ll{R_H}$) since it is only at this point
(or within ${k_B}T$ of it) that the charge on the puddle
can fluctuate. The maxima of the oscillations seen in
Willett {\it et al.}'s experiment~\cite{Willett08,Willett09a,Willett09u} would be
these peaks. If the density in the puddle is fixed,
then the spacing between peaks as a function of area
is naively just the additional area required to allow one more
electron into the puddle:
\begin{equation}
\label{eq:CB_odd}
\Delta A = \frac{e}{\rho_0}
\end{equation}
where $\rho_0$ is the charge density inside the dot.
However, in the case of a paired state, one would
expect that it is easier
to add an electron when the electron number is
odd than when it is even since, in the latter case,
an unpaired fermionic excitation is necessarily created.
So one would expect that, instead of evenly-spaced
peaks, the interval between an odd peak and the next even peak
would be smaller than the interval between an even peak
and the next odd peak because $V_s$ must also supply
the energy needed to create an unpaired fermionic excitation.
Consequently, the peak spacing would alternate between~\cite{Ilan08a}
\begin{equation}
\label{eq:CB_even}
\Delta A_{\pm} = \frac{e}{\rho_0}
\left( 1 \pm \frac{v_{n}}{2v_c} \right)
.
\end{equation}
As a result of this ``bunching'' effect, the periodicity
would be the interval between two successive even peaks,
i.e. twice what one might ordinarily expect.
But when there is an odd number of charge $e/4$ quasiparticles
in the MR, $\overline{\text{Pf}}$, or SU$(2)_2$ states, the
minimum energy to create a fermionic excitation is zero.
Thus, there is no bunching effect in this case, and
the period is not doubled~\cite{Stern06a}.

In the case of the $(3,3,1)$ state, bunching generically occurs with either an even or odd number of quasiparticles in the puddle.
However, when $n_q$ is odd, the bunching depends on the strength of the violation of $S_z$ conservation (where $S_z$ is the
$z$-component of the spin or, if one contemplates a bi-layer version of this experiment, the layer pseudospin)
so $v_{n} / 2 v_c$ in Eq.~(\ref{eq:CB_even}) is replaced by a different constant dependent on this violation. This is because the neutral sector of the edge theory of the $(3,3,1)$ state is a pair of Majorana fermions or, equivalently, a Dirac fermion (which can be bosonized, leading to the standard $K$-matrix description). If $S_z$ is conserved (or only weakly non-conserved), both Majorana fermions will have edge zero modes for odd $n_q$ and the bunching will disappear. In this case, the bunching pattern would look just like that predicted for the MR, $\overline{\text{Pf}}$, and SU$(2)_2$ states. However, if $S_z$ is more than weakly non-conserved, then the two Majorana fermions zero modes will mix and split, leaving no zero modes. Consequently, when $S_z$ is not conserved, there will be bunching even when there is an odd number of quasiparticles. Thus, switching between bunching and non-bunching regions in Coulomb blockade at $\nu = 5/2$ is not necessarily an indication of a non-Abelian state. Furthermore, if ${v_n} \ll {v_c}$, bunching will never be seen in any state. More generally, the switching between different bunching patterns in Coulomb blockade described in Refs.~\onlinecite{Stern06a,Ilan08a} for non-Abelian states may similarly be mimicked by corresponding Abelian states (for more details, see Ref.~\onlinecite{Bonderson09d} where explicit examples are given for all the most physically relevant non-Abelian states).

The strongly-paired $K=8$ state~\cite{Halperin83} always exhibits bunching, now with $v_{n} / 2 v_c$ in Eq.~(\ref{eq:CB_even}) replaced by a constant dependent upon the finite energy cost of having an unpaired electron. If this energy cost is small, it may not appear bunched. On the other hand, if it is large enough, it will be maximally bunched with $\Delta A = 2e / \rho_0$ corresponding to tunneling electron pairs.

\subsection{Non-linear Area vs. $V_{s}$ dependence}

The assumptions that $\Delta A=c\Delta V_s$ with only a single value of $c$ across
a range of filling factors and a range of $V_s$ values are important for the two previous interpretations of this experiment.
It is not clear that $dA/d{V_s}$ should be constant across an appreciable range
of $V_s$ values because the density is not constant across
the device.
In fact, we expect $V_s$ to vary linearly with
total charge in the central puddle. So long as the electron density is
essentially fixed, apart from a small number of quasiparticles,
$V_s$ will vary linearly with $A$. However, if there are high-density
and low-density regions, then we will have
$\Delta A=c\Delta V_s$ in some regions and
$\Delta A=c'\Delta V_s$ in the others, with $c\neq c'$.
(This could lead, for instance, to $\nu=7/3$ puddles
within the $\nu=5/2$ droplet.)
This would, in turn, lead to two different
regions with different oscillation periods.
However, it is difficult to see why one period would be twice
the other or why there would be two periods only at
$\nu=5/2$ and not at $\nu=5/3,2,7/3$.

One other possibility, which also depends on spatial inhomogeneity
although still assuming a linear $A$ vs $V_s$, is that there
are regions in the sample in which the $K=8$ Abelian state occurs.
The rest of the state is assumed to be non-Abelian, i.e. either the MR or $\overline{\text{Pf}}$
state. Then, when a $K=8$ region is at the edge of the system,
varying $V_s$ does not change the area enclosed by the edge of the
non-Abelian part of the system, which would lead to $e/4$ oscillations.
It does cause the total area to vary, but this only causes $e/2$
oscillations since these oscillations can move coherently along
both $K=8$ Abelian and non-Abelian edges. Thus, the two regions
correspond to when the edge of the system near the side gate is
a $K=8$ region or a non-Abelian region.

\section{Critical Analysis}
\label{sec:critical}

\subsection{Non-Abelian Interference}

The data of Refs.~\onlinecite{Willett08,Willett09a,Willett09u} are broadly consistent with the hypothesis that the device is functioning
as a quantum Hall edge state interferometer. As
the temperature is raised, the putative $e/4$
oscillations, which are observed at 30~mK,
disappear while the $e/2$ oscillations persist
even at 150~mK. This was anticipated in Ref.~\onlinecite{Wan08a},
where it was noted that the coherence length
will be substantially longer for $e/2$ quasiparticles than
for $e/4$ quasiparticles since the former do not
involve the slow neutral edge modes.
 Thus, any of the proposed $\nu=5/2$ states
(apart from the strong-pairing state)
would be broadly consistent with the $e/4$ oscillations
seen in Refs.~\onlinecite{Willett08,Willett09a,Willett09u}.
However, there is no simple explanation of their
absence in the type II regions in the $(3,3,1)$
state, while the MR, $\overline{\text{Pf}}$, and SU(2)$_2$
states all provide a simple explanation, as described in
the previous section. One however needs a more careful and detailed study of the temperature and voltage behavior before a favored candidate non-Abelian state can be identified.

Perhaps the most serious challenge to the non-Abelian
interferometer hypothesis is that the same amplitude of $e/2$ oscillations should
always be present (i.e. in Type I and II regions), while $e/4$ oscillations should only be
observed when the quasiparticle number contained within the interferometry region is even. This appears to be the
case in Figs.~3 and S2b of Ref.~\onlinecite{Willett09a} and Fig.~2a of Ref.~\onlinecite{Willett09u}, and to a lesser extent in Figs.~4b and S2a of Ref.~\onlinecite{Willett09a} and Fig.~2c of Ref.~\onlinecite{Willett09u}. However, this appears not to be the case in Fig.~4a of Ref.~\onlinecite{Willett09a} and Fig.~2b of Ref.~\onlinecite{Willett09u}. It is possible to generate some accidental destructive interference between the oscillations due to tunneling of $e/2$ quasiparticles given in Eq.~(\ref{eq:I12e2}) and that of double pass interference of $e/4$ quasiparticles given in Eq.~(\ref{eq:I1212e4}), since the relative phase of these terms is not fixed. This could potentially result in the appearance and disappearance of $e/2$ oscillations, however, as previously mentioned, the magnitude of oscillations in Eq.~(\ref{eq:I12e2}) is so strongly suppressed in the experiments of Refs.~\onlinecite{Willett08,Willett09a,Willett09u} that it could not explain such behavior there.

As we describe in Appendix~\ref{sec:matrix-elements},
a simple model of quasiparticle tunneling predicts
that the amplitude for $e/4$ quasiparticle
backscattering, $\Gamma_{e/4}$, is much larger than the
amplitude for $e/2$ quasiparticle
backscattering, $\Gamma_{e/2}$.
However, the magnitude of $e/2$ oscillations
in the type II regions is comparable to the magnitude of $e/4$ oscillations in the type I regions.
It may be that $\left| \Gamma_{e/2} \right|$
is ``accidentally'' large, e.g. due to the presence of a tunneling resonance for $e/2$ quasiparticles at the point contact. (For example, such a resonance might occur if there is an impurity or a region of different filling in the point contact acting like a dot or anti-dot through which resonant tunneling favors the $e/2$ charge.)
Alternatively, as a result of the shorter coherence length for $e/4$
excitations, the corresponding oscillations
are more strongly suppressed. This would
require a coincidence -- that thermal smearing
of $e/4$ excitations compensates for the smallness of the ratio $\Gamma_{e/2}/\Gamma_{e/4}$.
However, this could be tested by going to lower temperatures to reduce the suppression, and by increasing the separation between the point contacts to increase the suppression.
At any rate, given that $e/2$ oscillations are observed
in the type II regions, it would be a problem for the
non-Abelian interferometer picture if they are not
generically seen in the type I regions.

However, it is worth noting in this context that the presence of
charge $e/2$ quasiparticle tunneling is not manifest
in the point contact experiments of Refs.~\onlinecite{Dolev08,Radu08}. In the former, the shot noise
appears to indicate that only charge $e/4$ quasiparticles
tunnel at the point contact (although there is sufficient
scatter in the data that one might argue that there could
be a component due to $e/2$ quasiparticle, the scatter
does not seem to be asymmetric in the direction of
charges larger than $e/4$ as one might have expected).
In the latter experiment, the best fit to the data
is actually ${e^*}/e=0.17$, so including any $e/2$ tunneling leads to a worse fit to the data~\cite{Radu08}.
Thus, the appearance and strength of $e/2$ quasiparticle tunneling remains
a mystery in several different experiments.

\subsection{Coulomb Blockade}

A conventional Coulomb blockade picture seems
inappropriate since $I_b \lesssim .1 I_{\text{total}}$
indicates that the system is in the weak back-scattering limit.
It is also unlikely that Coulomb blockade could
lead to two distinguishable periods since,
for $v_{n} / 2 v_c$ small (as we expect it to be),
the bunching will be difficult to resolve. Numerical
calculations of the edge velocities~\cite{Wan08a}
give $v_{n} \simeq 0.1 v_c$, confirming this expectation.
On the other hand, we note that Coulomb blockade is capable of producing peaks that alternate between the $e/4$ and $e/2$ periodicities, with no $e/2$ background in the $e/4$ region. Thus, if the two prior points against it were somehow incorrect, Coulomb blockade could provide a consistent explanation of the periodicity issue.

Furthermore, Coulomb blockade could be easily ruled out by measuring its temperature dependence and its dependence on
asymmetry between the tunneling amplitudes at the two point contacts. In particular, the Coulomb blockade peak widths are expected to scale linearly with temperature~\cite{Folk96}.
However, a more general view of Coulomb
blockade has emerged~\cite{Zhang09} (see, also
Refs.~\onlinecite{Rosenow07,Ofek09}), according to
which Coulomb blockade (CB) can be distinguished from
AB interference by {\it inter alia}
the dependence of $\Delta V_s$ on $B$
(it should be inversely proportional
for AB and independent for CB). This more general
view of Coulomb blockade is probably better described
as ``Coulomb dominated'' since it corresponds to a regime
in which the charging energy of the puddle between the
point contacts is the dominant energy scale. It does {\it not}
rule out a simple interpretation of the backscattered current
according to Eq.~(\ref{eq:nA_int_5_2}).

At any rate, by this criterion as
well, the data appears to be more consistent with
AB interference since $(\Delta {V_s})_{5/3}B_{5/3}\approx
(\Delta {V_s})_{7/3}B_{7/3}\approx 3\cdot (\Delta {V_s})_{2}B_{2}$.
However, it is worth keeping in mind
that we do not know precisely how the area of the
droplet changes with $V_s$ or with $B$ (the $B$ dependence further distinguishes
Aharonov-Bohm from Coulomb blockade because $A$ should be independent
of $B$ in the former case, but not in the latter case); knowing
this would enable us to cement an interpretation of the
experiment.

\subsection{Non-linear Area vs. $V_s$ dependence}

As previously mentioned, one might question the validity of the assumption that $\Delta A={c}\Delta V_s$ holds with the same value of $c$ across a range of filling fractions. However, the assumption that
$c$ is independent of the filling fraction for nearby filling fractions
is, in fact, reasonable.
$V_s$ is several volts, and the oscillation periods are
$\sim 10$~mV which are much higher energy scales than
the weak energy gaps and correlation effects associated with
the $\nu=5/3,7/3,5/2$ quantum Hall states. Thus, the details of
these quantum Hall states are probably unimportant and
$dA/d{V_s}$ is probably determined by the electric potential
due to the donor impurities and the electron density, which are not
varying significantly. However, when there are filled Landau
levels beneath the quantum Hall state of interest,
their edges can screen the side-gate voltage, presumably
weakening the dependence of $A$ on $V_s$ (since $A$
is the area of the droplet of the fractional state in
the partially-filled Landau level). In particular,
we would expect $\Delta A={c_1}\Delta V_s$ at
$\nu=1/3$ but $\Delta A={c_2}\Delta V_s$ at $\nu=7/3$,
with ${c_1}>{c_2}$. However, by the same reasoning,
we expect that the relationship between $A$ and $V_s$
will be the same for $\nu=5/3,7/3,5/2$ (if the $\nu=5/3$
edge is two filled Landau levels with a backward
propagating $\nu=1/3$ edge mode).

\section{Additional proposals and non-trivial checks}
\label{sec:proposals}

As beautiful as the non-Abelian anyon explanation
of the results of Ref.~\onlinecite{Willett08,Willett09a,Willett09u} may be,
it is clear from the preceding analysis
that there are some significant gaps
which need to be closed through further measurements.

\subsection{Temperature and voltage scaling behavior}

If it is, indeed, the case that $R_L$ is due to the
weak backscattering of $e/4$ quasiparticles
at the constrictions, then both the
non-oscillatory and oscillatory parts of the current
should have non-trivial temperature and voltage
dependence. Modeling the edge in the simplest way (i.e. fully equilibrated neutral modes and no edge reconstruction) using the ``natural'' conformal field theory inherited from the bulk, one can perform a more detailed analysis of the tunneling edge
current~\cite{Fendley06a,Fendley07a,Lee07,Levin07},
along the lines of that carried out in Refs.~\onlinecite{Wen92b,Chamon97} for Abelian states.

The non-oscillatory part of the backscattered current --
the sum of the contributions from each point
contact independently -- will behave as the power laws:
\begin{equation}
\label{eqn:power-laws}
I_{b}^{\left(qp\right)} \propto \left\{
\begin{array}{ll}
T^{2g -2}\,V  & \text{for small } eV \ll {k_B}T \\
V^{2g - 1}    & \text{for small } eV\gg {k_B}T
\end{array}
\right.
,
\end{equation}
where $g = g_{c} + g_{n}$
is the tunneling exponent combining charge and neutral (Abelian and non-Abelian) sectors of the quasiparticles' tunneling operator. The tunneling operator is relevant for $g<1$, and quasiparticles with smaller $g$ are more relevant, and are thus expected to dominate the tunneling current in the weak backscattering limit.

\begin{table}
\[
\begin{array}{|l|c|c|c|c|c|c|}
\hline
\nu = \frac{5}{2}            &   e^{\ast}   &   \text{n-A?}  &   \theta          &   g_c   &   g_n   &   g    \\
\hline
\text{MR:}                   &   e/4        &   \text{yes}   &   e^{i \pi /4}    &   1/8   &   1/8   &   1/4  \\
                             &   e/2        &   \text{no}    &   e^{i \pi /2}    &   1/2   &   0     &   1/2  \\
\hline
\overline{\text{Pf}}\text{:} &   e/4        &   \text{yes}   &   e^{-i \pi /4}   &   1/8   &   3/8   &   1/2  \\
                             &   e/2        &   \text{no}    &   e^{i \pi /2}    &   1/2   &   0     &   1/2  \\
\hline
\text{SU$\left(2\right)_2$:} &   e/4        &   \text{yes}   &   e^{i \pi /2}    &   1/8   &   3/8   &   1/2  \\
                             &   e/2        &   \text{no}    &   e^{i \pi /2}    &   1/2   &   0     &   1/2  \\
\hline
\text{K=8:}                  &   e/4        &   \text{no}    &   e^{i \pi /8}    &   1/8   &   0     &   1/8  \\
                             &   e/2        &   \text{no}    &   e^{i \pi /2}    &   1/2   &   0     &   1/2  \\
\hline
\text{(3,3,1):}              &   e/4        &   \text{no}    &   e^{i 3\pi /8}   &   1/8   &   1/4   &   3/8  \\
                             &   e/2        &   \text{no}    &   e^{i \pi /2}    &   1/2   &   0     &   1/2  \\
\hline
\end{array}
\]
\caption{Relevant quasiparticle excitations of model FQH states at $\nu = 5/2$. Here we list their values of charge $e^{\ast}$; whether they are non-Abelian; their topological twist factor $\theta$; and their charge and neutral scaling exponents $g_c$, $g_n$, and $g$. The MR, $\overline{\text{Pf}}$, and SU$(2)_2$ NAF states are non-Abelian, while the K=8 (strong pairing) and (3,3,1) states are Abelian. All of these have Abelian $e/2$ Laughlin-type quasiparticles.}
\label{Table:qps}
\end{table}

From Table~\ref{Table:qps}, we see that
the $e/4$ backscattering operator is a relevant
perturbation of the edge effective theory
for all of the candidate states.
Thus, the effective tunneling amplitude(s) will
decrease as the temperature is raised,
as $T^{-3/2}$, $T^{-5/4}$, or $T^{-1}$ in the MR, $(3,3,1)$,
or $\overline{\text{Pf}}$ and SU$(2)_2$ states, respectively.
Charge $e/2$ backscattering is also relevant in
all of the candidate states.
Because $e/2$ excitations have $g=1/2$ and are entirely
in the charge sector, their lowest order, single point-contact tunneling current
contribution is the same in all of the candidate states. Using the methods of Ref.~\onlinecite{Wen92b} to explicitly compute its value, we find \begin{equation}
I_{b}^{\left(e/2\right)} = \frac{e}{2}\left| \Gamma_{e/2} \right|^{2} \frac{2 \pi}{v_{c}} \tanh\left( \frac{e V}{4T} \right)
.
\end{equation}
Thus, in the linear response regime,
the effective tunneling amplitude for $e/2$
backscattering decreases as $T^{-1}$.
For the MR and Abelian states, charge $e/4$ backscattering is more relevant than charge $e/2$ backscattering, so it is expected to dominate at lower temperatures.
For the $\overline{\text{Pf}}$ and SU$(2)_2$ states, $e/4$ and $e/2$
backscattering are equally relevant (with $g=1/2$).
As previously mentioned in Section~\ref{sec:introduction}, the single point contact experiment of Ref.~\onlinecite{Radu08} found voltage and temperature dependence of the tunneling current to be most consistent with the $\overline{\text{Pf}}$ and SU$(2)_2$ states, providing a point of reference with which to compare.

Turning now to the oscillatory current, we note that
for $eV\ll {v_n}/2L$, where $2L$ is the interference
path length, it has the same voltage
dependence as the non-oscillatory current.
For larger voltages, it becomes apparent that there
are oscillations with a period $\sim 4\pi {v_n}/L$,
but these are much larger voltages than are probed
in the experiments of Refs.~\onlinecite{Willett08,Willett09a,Willett09u}
We note that these oscillations can be turned around and
interpreted as oscillations as a function of $L$,
which changes when the interferometry area is changed
(i.e. when the side gate voltage is changed).
However, these oscillations have periodicity $4 \pi v / e^{\ast} V$, where $v$ represents several characteristic velocities, which are all dominated by the slowest edge mode velocity (which is expected to be $v_n$). Since $V \simeq 10^{-8}$~V,
these will only give rise to envelopes with periods much longer than that of the oscillations observed in the experiment.

The temperature dependence of the oscillatory
current includes a power-law prefactor
of the form in Eq.~(\ref{eqn:power-laws}) in addition
to the exponential suppression
$e^{-T/T^{\ast}\!(L)}= e^{-L/L_{\phi}(T)}$
which we discussed earlier.
Thus, the relative suppression of the
$e/4$ contribution, compared to the $e/2$ contribution,
must be due entirely to the shorter coherence length
in the $\overline{\text{Pf}}$ case but could be
due to a combination of effects in the MR or $(3,3,1)$
case.

To make the case for interference stronger,
it would be helpful to disentangle the effects
of the temperature dependence of the coherence
length from the temperature dependence of the
effective tunneling amplitude. One way to do this
would be to carefully study the bias voltage
dependence of the current backscattered
by the interferometer of Refs.~\onlinecite{Willett08,Willett09a,Willett09u}
at some fixed $V_s$ in the low-$T$ limit.
If the behavior is similar to
that observed in Ref.~\onlinecite{Radu08} (and, especially,
if it is the behavior expected for one of the possible
$\nu=5/2$ states), then this is a strong indication
that $R_L$ is due to the weak backscattering of
charge $e/4$ quasiparticles. Another useful way to do this would be to turn on the point-contacts one at a time and study their tunneling behavior individually. This would help determine which state
occurs in the point-contact region; it is important that it is
at the same filling fraction as the rest of the bulk. Furthermore, it would allow one to determine the relative tunneling amplitudes of $e/4$ and $e/2$ quasiparticles and confirm that the experiment is not in the CB regime.

\subsection{Fourier analysis}

It is important to verify that the oscillation periodicities behave as expected. In addition to confirming the dominance of $\Delta A = 4 \Phi_0 / B$ and $\Delta A = 2 \Phi_0 / B$ oscillations in the Fourier spectrum, one should examine the spectrum in the different regions more carefully. Specifically, by using windowing techniques in the Fourier analysis of the data, one should check that the type I regions have both $\Delta A = 4 \Phi_0 / B$ and $\Delta A = 2 \Phi_0 / B$ oscillations, that the type II regions have only $\Delta A = 2 \Phi_0 / B$ oscillations, and that the amplitude of $\Delta A = 2 \Phi_0 / B$ oscillations are roughly the same in the type I and II regions. It is also useful to know the relative oscillations amplitudes of the two frequencies in the type I regions.

\subsection{Significance of $\pi/2$ and $\pi$ phase shifts}

Another aspect of the data worth examining more closely is the phase shifts observed in the oscillations of a given type. While an $e/4$ bulk quasiparticle entering or exiting the interference loop causes a switch between type I and II oscillations, there could also be $e/2$ quasiparticles in the bulk that enter or exit the interference loop when the area is changed (or pairs of $e/4$ quasiparticles that enter or exit nearly simultaneously), which would not switch the oscillations between type I and II. From Eq.~(\ref{eq:I12e2}) we see that this would cause a shift of phase $\pi$ in the $e/2$ oscillations, while from Eq.~(\ref{eq:nA_int_5_2}) we see that this would cause a shift of phase $\pi/2$ in the $e/4$ oscillations. To more firmly establish the origin of these phase shifts, as well as how reproducible are the fine details of the observed oscillations, one should immediately backtrack on the side gate voltage within an oscillation region of fixed type to see whether the oscillations (including locations of phase shifts) are nearly exactly reproduced. Furthermore, we recall that if one is in a type I region and an $e/4$ bulk quasiparticle exits and then re-enters the interference loop, the value of $n_{\psi}$ in Eq.~(\ref{eq:nA_int_5_2}) will be randomized with equal probability. Hence, one should also increase the side gate voltage (decreasing the interferometry area) until one transitions from a type I region into a type II region, and then soon afterward backtrack on the side gate voltage to return to the type I region. When this is done repeatedly, if the $e/4$ oscillations in the type I region are found to be shifted by a phase of $\pi$ half the time, it would be a direct observation of non-Abelian statistics. Doing this backtracking immediately would provide a more convincing observation of the $\pi$ phase shift due to the non-Abelian statistics than backtracking after a complete run, which leaves plenty of time (one day) for small changes to occur that may give non-universal phase shifts. Furthermore, it is more convenient to perform a statistical experiment for which each sampling takes one hour rather than one day to complete.

\subsection{Area of the interferometry region}

It is important to establish that the periodicities with $V_s$ obtained in the experiment correspond directly to periodicities with
$A$. One way to attempt to do this would be to
vary both $V_s$ and $B$ at $\nu=2$ and to
use the periodicity in $B$ to determine the area
for several different values of $V_s$. One could,
in this way, check that the assumed constant $c$ in
$\Delta A = c \Delta V_{s}$ is really constant. Such a
measurement would also determine whether the oscillation
pattern corresponds to AB interference or Coulomb blockade,
as in Ref.~\onlinecite{Zhang09}.
One could also check that the oscillations are due to
AB interference by turning down or off one of the point contacts
(and then the other) and repeating the experiment, which should
cause the oscillations to disappear. This will further exclude resonances at a single point contact as the source of oscillations and will give a better value of the tunneling amplitude
for a single point contact.

A more ambitious approach to measuring the area within
the interference loop, which could simultaneously tackle the even more fundamental
problem of determining directly whether the type I and II
regions correspond to even/odd quasiparticle numbers,
would be to image the two dimensional electron gas (2DEG) in the interferometer using a scanning single electron transistor (SET),
as in Ref.~\onlinecite{Martin04}. In this earlier experiment,
a scanning SET was used to image the compressibility of the electron liquid
at $\nu=1,1/3,2/3$. By measuring the compressibility,
it should be possible to determine where the edge of the
Hall fluid is in the droplet and, hence, the area of the interferometry region.
It should also be possible to find the localized states near
the Fermi energy where $e/4$ quasiparticles could be trapped.
By imaging the charge $e/4$ quasiparticles, one might even be able to see these
localized quasiparticles enter or leave the interference loop as $V_s$ is varied, and hence allow the most direct verification of the non-Abelian interferometer interpretation.

Similarly, the possibility that the existence of two
periodicities signals different regions with different
electron densities and, therefore, two different
possible relations $\Delta A = c \Delta V_{s}$ and
$\Delta A = c' \Delta V_{s}$ could be ruled in or
out through a more detailed knowledge of the electron
density in the sample.

\subsection{Tracking bulk quasiparticles}

We note, as a consistency check on the data of
Ref.~\onlinecite{Willett09a,Willett09u} that one can use the
area periodicity of the AB oscillations to estimate the density of bulk $e/4$ quasiparticles by attributing each observed switching between type I and II regions to an $e/4$ quasiparticle entering or leaving the interferometry region. In this way, we estimate the density to be $\rho_{e/4} \approx 50~\mu\text{m}^{-2}$. This translates to $\sim 10$ charge $e/4$ quasiparticles
in the interference loop. We can also estimate the number of $e/2$ quasiparticles in the bulk, though perhaps less reliably, by attributing phase disruptions observed within an oscillation type region to an $e/2$ quasiparticle crossing the interferometry region. Using this to similarly estimate the density of bulk $e/2$ quasiparticles gives $\rho_{e/2} \approx 50~\mu\text{m}^{-2}$, or roughly the same number as charge $e/4$ quasiparticles
in the interference loop. Depending on how seriously one takes the $e/2$ contribution, this gives approximately $0.5-1.5 \%$ depletion of the electron density in the bulk, which is the reported density variation
in the device~\cite{Willett08,Willett09a,Willett09u}.
With a scanning SET setup, it
may be possible to find the $\sim 10$ charge $e/4$ quasiparticles
that are necessary for the non-Abelian interferometer
interpretation and observe them entering or exiting the interference loop.
Localized $e/2$ quasiparticles or, equivalently,
closely-spaced pairs of $e/4$ quasiparticles should also be
observable.

\subsection{Multiple plungers}

A more crude, but also more easily implementable way to further strengthen the correlation between which oscillation type is observed and the localization of excitations in the bulk is to independently vary two or more plunger gates of the interferometer. By refining the ability to control how the interferometry area is changed beyond a single plunger variable, the changes between oscillation types can be more strongly associated with a particular area. If a region of one plunger's gate voltage exhibits type I oscillations, but then, after changing a separate plunger's position, the same voltage range in the first plunger exhibits type II oscillations, this would demonstrate that a particular oscillation type is not associated with that particular voltage range of the first plunger, but rather that an ability to change between oscillation types is associated with a localized quantity in the area added or removed by the second plunger. This would greatly strengthen the evidence for non-Abelian braiding statistics.

\subsection{Other second Landau level states}

A double point-contact interferometer may also be
used to test whether the quantum Hall states at $\nu=7/3$, $12/5$, $8/3$, and $14/5$
are non-Abelian. These filling fractions all have compelling Abelian alternatives which almost certainly occur
at their corresponding lowest Landau level counterparts. While numerical studies strongly support the MR and $\overline{\text{Pf}}$ states at $\nu=5/2$~\cite{Morf98,Rezayi00,Feiguin08a,Moller08a,Wan08a,Peterson08b,Feiguin08b,Peterson08c,Storni08} and the (particle-hole conjugate) Laughlin state at $\nu=14/5$~\cite{dAmbrumenil88,Peterson08a}, they are far less conclusive for $\nu=7/3$, $12/5$, and $8/3$~\cite{Read99,Rezayi09a,Peterson08a,Bonderson09a}, where several candidates seem plausible, including ones that are non-Abelian. It is clearly important to also test these FQH states experimentally. The details of the plausible candidates' experimental signatures are discussed in Appendix~\ref{sec:other-candidates}. The signatures of non-Abelian statistics in these states will again be dramatic, though not quite as
much as for the MR, $\overline{\text{Pf}}$, or SU$(2)_2$ NAF states.

\section{Discussion}
\label{sec:discussion}

We close this discussion by assuming, for a moment,
that the experiments of
Refs.~\onlinecite{Willett08,Willett09a,Willett09u} are,
in fact, performing interferometry on the
$\nu=5/2$ state of the sort envisioned in Refs.~\onlinecite{Chamon97,Fradkin98,DasSarma05,Stern06a,Bonderson06a}
and are detecting non-Abelian quasiparticles.
What forecast would these results give for topological
quantum computation~\cite{Kitaev03,Preskill98,Freedman98,Freedman02a,Freedman02b,Freedman03b,Preskill-lectures,Nayak08}?
Certainly, it would be encouraging that a non-Abelian
topological state, the {\it sine qua non} for topological
quantum computing, would be found. One potential
source of concern is the appearance of some
seemingly unpredictable phase disruptions, which would
make it difficult to distinguish the two states of a topological
qubit, which differ by a $\pi$ phase shift in their ($e/4$ oscillation) interference patterns. However, if further investigation shows that
they are $\pi/2$ phase shifts (in the type I regions), then they may be attributable to $e/2$ quasiparticles or pairs
of $e/4$ quasiparticles simultaneously entering or exiting the interference loop, and it would be a manageable problem. On the other hand, if they turn out to be $\pi$ phase shifts in the $e/4$ oscillations, then they could be attributed to either tunneling of a $e/4$ quasiparticle from one edge to the other between bulk quasiparticles within the interferometry region, or tunneling of an electrically neutral fermion between the edge and a bulk quasiparticle within the interferometry region. Either of these $\pi$ phase shifting processes would cause errors in topological qubits, so if we attribute all the observed phase disruptions to such processes, this gives a crude estimate of about an hour for the time scale for such errors, or a contribution to the topological qubit error rate of $\Gamma/\Delta \lesssim 10^{-13}$. To better determine the rate of such errors, one could simply tune the voltage to a local $e/4$ oscillation maximum/minimum of the tunneling current in a type I region and time average duration it takes for the current to jump to a lower/higher value (i.e. experiencing a $\pi$ phase shift to a minimum/maximum of the current). If the phase disruptions are neither $\pi/2$ nor $\pi$ phase shifts, then they would be a serious concern, as they would have no obvious explanation. Finally, the apparent stability of the type I and type II regions implies that thermally-activated charge $e/4$ quasiparticles do not move in and out of the interferometry region over the time scales of this experiment. Indeed, these regions are stable on a time scale of a week, which would imply a topological
qubit error rate from mobile bulk quasiparticles of $\Gamma/\Delta \leq 10^{-15}$.

``With luck, we might see a topological qubit within a year.'' -- attributed to Kirill Shtengel, January 9, 2009~\cite{PB-CN-private}.

\begin{acknowledgments}
We would like to thank R.~Willett for many important discussions and access to unpublished work, M. Heiblum for a discussion of unpublished work, and S.~Das~Sarma, M.~Freedman, and Y.~Gefen for helpful comments. WB, KS, and JKS would like to acknowledge the support and hospitality of Microsoft Station Q. PB, CN, and KS would like to acknowledge the hospitality of the Aspen Center for Physics. CN and KS are supported in part by the DARPA-QuEST program. KS is supported in part by the NSF under grants DMR-0748925 and PHY05-51164, and would like to acknowledge the hospitality of the KITP. JKS is supported in part by the Science Foundation of Ireland Principal Investigator grant 08/IN.1/I1961.
\end{acknowledgments}

\appendix
\section{Other second Landau level states}
\label{sec:other-candidates}

In this section, we consider the other observed FQH states in the second Landau level~\cite{Xia04,Pan08}, i.e. $\nu = 7/3$, $12/5$, $8/3$, and $14/5$. We provide the data of their prominent candidate descriptions that will be useful for interpreting tunneling and interference experiments, similar to earlier in this paper.

As described earlier, the interference term of the tunneling current combines: (1) the AB effect, (2) the braiding statistics with encircled quasiparticles, and (3) the edge physics. The AB effect simply contributes a phase $e^{ i e^{\ast} \Phi /\hbar c} = e^{ i 2\pi e^{\ast} \Phi / e \Phi_0}$ when the quasiparticle of charge $e^{\ast}$ encircles flux $\Phi$. The physics describing propagation of excitations on the edge gives rise to a temperature, bias voltage, and interference path length dependence of the tunneling edge current~\cite{Wen92b,Chamon97,Bishara08} that we denote as $F\left( T,V,L \right)$. The details of this edge physics can generally be complicated, but the most significant aspect is the coherence length and temperature, which is given as before in Eqs.~(\ref{eq:coh_length},\ref{eq:coh_temp}), with the appropriate scaling exponents for excitations of a given state (which are given in the tables).

For the lowest order tunneling interference process, the braiding statistics contributes the factor~\cite{Bonderson06b}
\begin{equation}
M_{ab} = \frac{S_{ab} S_{00}}{S_{0a} S_{0b}}
,
\end{equation}
where $S_{ab}$ is the topological $S$-matrix, and $a$ and $b$ are the topological charges of the tunneling edge excitation and the encircled bulk quasiparticle excitations, respectively. These combine to give the lowest order interference contribution to the tunneling current (in the asymptotic limit where the state of the bulk quasiparticles is projected onto a definite value of $b$)
\begin{equation}
\label{eq:nA_int_gen}
I^{\left( qp \right)}_{12} \propto \text{Re} \left\{ e^{i 2\pi \frac{ e^{\ast} \Phi }{e \Phi_0} } \, M_{ab}^{\ast} \,\, F\left( T,V,L \right) \right\}
.
\end{equation}
If either $a$ or $b$ is an Abelian charge, $M_{ab}$ is simply a phase. More generally, when $a$ and $b$ are both non-Abelian charges, $M_{ab}$ is a complex number with $\left| M_{ab} \right| \leq 1$. This leads to the potential for a suppression of the interference term [Eq.~(\ref{eq:nA_int_gen})] resulting from non-Abelian braiding statistics, similar to the non-Abelian $\nu=5/2$ states.

For the non-Abelian FQH states considered here, the braiding statistics are essentially given by the SU$(2)_k$ theories~\cite{Witten89}, up to Abelian phase factors. These theories have topological charges $j=0,1/2,1,\ldots,k/2$ and
\begin{equation}
M_{j_{1} j_{2}} = \frac{ \sin\left( \frac{ \left( 2j_{1}+1 \right) \left( 2j_{2}+1 \right) \pi }{k+2} \right) \sin\left( \frac{ \pi }{k+2} \right)}{ \sin\left( \frac{ \left( 2j_{1}+1 \right) \pi }{k+2} \right) \sin\left( \frac{ \left( 2j_{2}+1 \right)\pi }{k+2} \right)}
.
\end{equation}
The $k=2$, $3$, and $4$ cases are the most pertinent to our discussion, so we write them out explicitly:
\begin{equation}
\label{eq:M2}
M^{(2)} = \left[
\begin{array}{rrr}
1   &   1   &   1   \\
1   &   0   &  -1   \\
1   &  -1   &   1
\end{array}
\right]
,
\end{equation}
\begin{equation}
\label{eq:M3}
M^{(3)}= \left[
\begin{array}{cccr}
1   &    1           &   1           &   1   \\
1   &    \phi^{-2}   &  -\phi^{-2}   &  -1   \\
1   &   -\phi^{-2}   &  -\phi^{-2}   &   1   \\
1   &   -1           &   1           &  -1
\end{array}
\right]
,
\end{equation}
where $\phi = \frac{1 + \sqrt{5}}{2}$ is the Golden ratio, and
\begin{equation}
\label{eq:M4}
M^{(4)} = \left[
\begin{array}{rcrcr}
1   &   1                    &   1             &   1                    &   1   \\
1   &   \frac{1}{\sqrt{3}}   &   0             &  \frac{-1}{\sqrt{3}}   &  -1   \\
1   &   0                    &   \frac{1}{2}   &   0                    &   1   \\
1   &  \frac{-1}{\sqrt{3}}   &   0             &   \frac{1}{\sqrt{3}}   &  -1   \\
1   &  -1                    &   1             &  -1                    &   1
\end{array}
\right]
.
\end{equation}

The braiding statistics of the MR, $\overline{\text{Pf}}$, and SU$(2)_2$ NAF $\nu =5/2$ states are all derived from SU$(2)_2$. The non-Abelian quasiparticles in these states carry SU$(2)_2$ charge $1/2$. It follows that an odd number cluster of such quasiparticles will also carry a collective SU$(2)_2$ charge of $1/2$, while an even number cluster will carry either $0$ or $1$. Thus, looking at the ($j=1/2$) middle column of Eq.~(\ref{eq:M2}), we see exactly the source of the behavior described in Eq.~(\ref{eq:nA_int_5_2}).

\subsection{$\nu=7/3$}

For the $\nu=7/3$ FQH plateau, the leading candidates are the Laughlin (L) state~\cite{Laughlin83}, two types of Bonderson-Slingerland (BS) states~\cite{Bonderson07d}, and a $4$-clustered Read-Rezayi (RR) state~\cite{Read99}. (The bar indicates particle-hole conjugation.) The BS states considered here are hierarchically constructed over the MR and $\overline{\text{Pf}}$ states, and so have similar non-Abelian statistics derived from Eq.~(\ref{eq:M2}) using the fact that the non-Abelian quasiparticles carry SU$(2)_2$ charge $1/2$. The RR$_{k=4}$ state is related to SU$(2)_4$, and so has more complicated non-Abelian statistics, derived from Eq.~(\ref{eq:M4}). Its fundamental $e/6$ quasiparticles carry SU$(2)_4$ charge $1/2$.

\begin{table}[t!]
\[
\begin{array}{|l|c|c|c|c|c|c|}
\hline
\nu = \frac{7}{3}                   &   e^{\ast}   &   \text{n-A?}  &   \theta          &   g_c   &   g_n   &   g     \\
\hline
\text{L}_{1/3}\text{:}              &   e/3        &   \text{no}    &   e^{i \pi /3}    &   1/3   &   0     &   1/3   \\
\hline
\overline{\text{BS}}_{2/3}\text{:}  &   e/3        &   \text{yes}   &  e^{-i7\pi /24}   &   1/3   &   5/8   &   23/24 \\
                                    &   e/3        &   \text{no}    &  e^{i \pi /3}     &   1/3   &   0     &   1/3   \\
\hline
\text{BS}_{1/3}^{\psi}\text{:}      &   e/3        &   \text{yes}   &   e^{i 5\pi /24}  &   1/3   &   3/8   &   17/24 \\
                                    &   e/3        &   \text{no}    &   e^{i \pi /3}    &   1/3   &   0     &   1/3   \\
\hline
\overline{\text{RR}}_{k=4}\text{:}  &   e/6        &   \text{yes}   &   e^{-i \pi/6}    &   1/12  &   1/4   &   1/3   \\
                                    &   e/3        &   \text{no}    &   e^{i \pi /3}    &   1/3   &   0     &   1/3   \\
                                    &   e/2        &   \text{yes}   &   e^{i \pi /2}    &   3/4   &   1/4   &   1     \\
\hline
\end{array}
\]
\caption{Relevant quasiparticle excitations of model FQH states at $\nu = 7/3$. Here we list their values of charge $e^{\ast}$; whether they are non-Abelian; their topological twist factor $\theta$; and their charge and neutral scaling exponents $g_c$, $g_n$, and $g$. The $\overline{\text{BS}}$, BS$^{\psi}$, and $\overline{\text{RR}}_{k=4}$ states are non-Abelian, while the L state is Abelian. All of these have Abelian $e/3$ Laughlin-type quasiparticles. (Note: The $e/2$ excitation for $\overline{\text{RR}}$ is marginal, but we include it for the sake of representing the possibility of $e/2$ charge.)}
\label{Table_7_3:qps}
\end{table}

We see in Table~\ref{Table_7_3:qps} that all of these states have an $e/3$ excitation with smallest scaling exponent $g=1/3$, and so one expects these to dominate the tunneling. The $\overline{\text{RR}}_{k=4}$ state also has $e/6$ excitations with $g=1/3$, which should give a comparable contribution to the tunneling current. The experiments of Refs.~\onlinecite{Dolev08,Willett09a}, which appear to observe $e/3$ tunneling, but not $e/6$ tunneling at $\nu=7/3$, seem to exclude the $\overline{\text{RR}}_{k=4}$ state, while they agree with the L$_{1/3}$, $\overline{\text{BS}}_{2/3}$, and BS$^{\psi}_{1/3}$ states. In fact, since the relevant excitations of these latter three states all have $e/3$, and furthermore, the most relevant tunnelers are all Abelian, it will likely be difficult to distinguish between L$_{1/3}$, $\overline{\text{BS}}_{2/3}$, and BS$^{\psi}_{1/3}$ using tunneling and interferometry experiments. Thermal transport experiments are probably the best hope of distinguishing between these.

\subsection{$\nu = 12/5$}

For the $\nu=12/5$ FQH plateau, the leading candidates are the Haldane-Halperin (HH) state~\cite{Haldane83,Halperin84}, two types of BS states~\cite{Bonderson07d}, and a $3$-clustered RR state~\cite{Read99}. These BS states again have non-Abelian statistics derived from Eq.~(\ref{eq:M2}) using the fact that the non-Abelian quasiparticles carry SU$(2)_2$ charge $1/2$. The RR$_{k=3}$ state is related to SU$(2)_3$, and so has non-Abelian statistics derived from Eq.~(\ref{eq:M3}). Its fundamental $e/5$ quasiparticles carry SU$(2)_3$ charge $1/2$.

\begin{table}[t!]
\[
\begin{array}{|l|c|c|c|c|c|c|}
\hline
\nu = \frac{12}{5}                         &   e^{\ast}   &   \text{n-A?}  &   \theta          &   g_c   &   g_n    &   g     \\
\hline
\text{HH}_{2/5}\text{:}                    &   e/5        &   \text{no}    &   e^{i 3\pi /5}   &   1/5   &   2/5    &   3/5   \\
                                           &   2e/5       &   \text{no}    &   e^{i 2\pi /5}   &   2/5   &   0      &   2/5   \\
\hline
\text{BS}_{2/5}\text{:}                    &   e/5        &   \text{yes}   &   e^{i 9\pi /40}  &   1/10  &   1/8    &   9/40  \\
                                           &   e/5        &   \text{no}    &   e^{-i 2\pi/5}   &   1/10  &   1/2    &   3/5   \\
                                           &   2e/5       &   \text{no}    &   e^{i 2\pi /5}   &   2/5   &   0      &   2/5   \\
\hline
\overline{\text{BS}}_{3/5}^{\psi}\text{:}  &   e/5        &   \text{yes}   &   e^{-i11\pi/40}  &   1/10  &   3/8    &   19/40 \\
                                           &   e/5        &   \text{no}    &   e^{-i 2\pi/5}   &   1/10  &   1/2    &   3/5   \\
                                           &   2e/5       &   \text{no}    &   e^{i 2\pi /5}   &   2/5   &   0      &   2/5   \\
\hline
\overline{\text{RR}}_{k=3}\text{:}         &   e/5        &   \text{yes}   &   e^{-i \pi/5}    &   1/10  &   3/10   &   2/5   \\
                                           &   2e/5       &   \text{no}    &   e^{i 2\pi /5}   &   2/5   &   0      &   2/5   \\
\hline
\end{array}
\]
\caption{Relevant quasiparticle excitations of model FQH states at $\nu = 12/5$. Here we list their values of charge $e^{\ast}$; whether they are non-Abelian; their topological twist factor $\theta$; and their charge and neutral scaling exponents $g_c$, $g_n$, and $g$. The BS, $\overline{\text{BS}}^{\psi}$, and $\overline{\text{RR}}_{k=3}$ states are non-Abelian, while the HH state is Abelian. All of these have Abelian $2e/5$ Laughlin-type quasiparticles; all of these except $\overline{\text{RR}}_{k=3}$ have a relevant Abelian $e/5$ quasiparticle.}
\label{Table_12_5:qps}
\end{table}

We see in Table~\ref{Table_12_5:qps} that all of these states have an Abelian $2e/5$ excitation with scaling exponent $g=2/5$, so there should always be a background of such excitations in tunneling. The HH$_{2/5}$, BS$_{2/5}$, and $\overline{\text{BS}}_{3/5}^{\psi}$ states all have an Abelian $e/5$ excitation with $g=3/5$, so there should be a weaker background of these excitations in the tunneling. The smallest scaling exponent for the BS$_{2/5}$ state belongs to the non-Abelian $e/5$ excitation, which is therefore expected to dominate the tunneling in this state. The $\overline{\text{BS}}_{3/5}^{\psi}$ state has a non-Abelian $e/5$ excitation which has slightly less relevant tunneling operator than the $2e/5$ excitation. The $\overline{\text{RR}}_{k=3}$ state has a non-Abelian $e/5$ excitation with the same scaling exponent $g=2/5$ as the $2e/5$ excitation, so they should have roughly equal contribution to tunneling.

In interferometry experiments, the BS$_{2/5}$, $\overline{\text{BS}}_{3/5}^{\psi}$, and $\overline{\text{RR}}_{k=3}$ states will all exhibit $e/5$ oscillations that will sometimes be suppressed. However, there are important distinctions within this behavior that can distinguish between them. In particular, the BS states will exhibit an even-odd effect similar to Eq.~(\ref{eq:nA_int_5_2}), always returning to suppression for $n_{q_{0}}$ odd, where $n_{q_{0}}$ is the number of non-Abelian $e/5$ fundamental quasiparticles. On the other hand, the RR state can exhibit both suppression and full amplitude oscillations for all values of $n_q$, and it has a probability of switching between them when a given quasiparticle is taken in and out of the interferometry region. Furthermore, when the oscillations are suppressed for the BS state, the smaller amplitude $e/5$ oscillations will be due to tunneling of the Abelian $e/5$ excitations (which will always be present), because the non-Abelian excitation will have fully suppressed interference. The relative contribution to the tunneling of these excitations is not a fixed amount, and will change depending on temperature and voltage (i.e. they have different scaling). In contrast to this, the suppression that would be observed in the RR state is due entirely to the braiding statistics of the non-Abelian $e/5$ excitation, and the suppressed oscillation amplitude should always be a constant factor of $\phi^{-2} \approx 0.38$ smaller than the full oscillation amplitude.

\subsection{$\nu = 8/3$}

The candidates for the $\nu=8/3$ FQH plateau are, of course, similar to the $\nu=7/3$ candidates, since the filling fractions are particle-hole dual. We stress, however, that the physically observed states at these filling fractions need not be particle-hole dual to each other, since physical effects, such as Landau level mixing, will tend to break particle-hole symmetry at these fillings. The leading candidates are the Laughlin state~\cite{Laughlin83}, two types of BS states~\cite{Bonderson07d}, and a $4$-clustered RR state~\cite{Read99}. These BS states again have non-Abelian statistics derived from Eq.~(\ref{eq:M2}) using the fact that the non-Abelian quasiparticles carry SU$(2)_2$ charge $1/2$. The RR$_{k=4}$ state again is related to SU$(2)_4$, with non-Abelian statistics derived from Eq.~(\ref{eq:M4}) and fundamental $e/6$ quasiparticles carrying SU$(2)_4$ charge $1/2$.

\begin{table}[t!]
\[
\begin{array}{|l|c|c|c|c|c|c|}
\hline
\nu = \frac{8}{3}                          &   e^{\ast}   &   \text{n-A?}  &   \theta          &   g_c   &   g_n    &   g     \\
\hline
\overline{\text{L}}_{1/3}\text{:}          &   e/3        &   \text{no}    &   e^{-i \pi /3}   &   1/3   &   1/3    &   2/3   \\
                                           &   2e/3       &   \text{no}    &   e^{i 2\pi /3}   &   2/3   &   0      &   2/3   \\
\hline
\text{BS}_{2/3}\text{:}                    &   e/3        &   \text{yes}   &   e^{i 7\pi /24}  &   1/6   &   1/8    &   7/24  \\
                                           &   e/3        &   \text{no}    &   e^{i 2\pi /3}   &   1/3   &   1/3    &   2/3   \\
                                           &   2e/3       &   \text{no}    &   e^{i 2\pi /3}   &   2/3   &   0      &   2/3   \\
\hline
\overline{\text{BS}}_{1/3}^{\psi}\text{:}  &   e/3        &   \text{yes}   &   e^{-i 5\pi/24}  &   1/6   &   3/8    &   13/24 \\
                                           &   e/3        &   \text{no}    &   e^{i 2\pi /3}   &   1/6   &   1/2    &   2/3   \\
                                           &   2e/3       &   \text{no}    &   e^{i 2\pi /3}   &   2/3   &   0      &   2/3   \\
\hline
\text{RR}_{k=4}\text{:}                    &   e/6        &   \text{yes}   &   e^{i \pi /6}    &   1/24  &   1/8    &   1/6   \\
                                           &   e/3        &   \text{yes}   &   e^{i \pi /3}    &   1/6   &   1/6    &   1/3   \\
                                           &   e/2        &   \text{yes}   &   e^{i \pi /2}    &   3/8   &   1/8    &   1/2   \\
                                           &   2e/3       &   \text{no}    &   e^{i 2\pi /3}   &   2/3   &   0      &   2/3   \\
\hline
\end{array}
\]
\caption{Relevant quasiparticle excitations of model FQH states at $\nu = 8/3$. Here we list their values of charge $e^{\ast}$; whether they are non-Abelian; their topological twist factor $\theta$; and their charge and neutral scaling exponents $g_c$, $g_n$, and $g$. The BS, $\overline{\text{BS}}^{\psi}$, and $\text{RR}_{k=4}$ states are non-Abelian, while the $\overline{\text{L}}$ state is Abelian. All of these have Abelian $2e/3$ Laughlin-type quasiparticles; all of these except RR have a relevant Abelian $e/3$ quasiparticle.}
\label{Table_8_3:qps}
\end{table}

We see in Table~\ref{Table_8_3:qps} that all of these states have an Abelian $2e/3$ excitation with scaling exponent $g=2/3$, so there should always be a background of such excitations in tunneling. The $\overline{\text{L}}_{1/3}$, BS$_{2/3}$, and $\overline{\text{BS}}_{1/3}^{\psi}$ states all have an $e/3$ excitation also with $g=2/3$, so these two excitations are expected to have roughly equal contribution to the tunneling in these theories. However, the smallest scaling exponent for the BS$_{2/3}$ and $\overline{\text{BS}}_{1/3}^{\psi}$ states belong to non-Abelian $e/3$ excitations, which are therefore expected to dominate the tunneling in these states. The smallest scaling exponent for the RR$_{k=4}$ state belongs to the non-Abelian $e/6$ excitation, which should thus dominate tunneling. There are additional relevant tunnelers for RR$_{k=4}$ that are non-Abelian with different statistics than the fundamental quasiparticle, namely the $e/3$ and $e/2$ excitations which carry SU$(2)_4$ charges $1$ and $3/2$, respectively. The experiments of Ref.~\onlinecite{Dolev08}, which observes only $e/3$ tunneling, appear to exclude the RR$_{k=4}$ state and best agree with the BS$_{2/3}$ and $\overline{\text{BS}}_{1/3}^{\psi}$ states.

\subsection{$\nu = 14/5$}

The $\nu=14/5$ FQH plateau is most likely the standard (particle-hole conjugate) Laughlin state~\cite{Laughlin83}, but we include this filling fraction for completeness, and list BS states~\cite{Bonderson07d} as (unlikely) alternative candidates. These BS states again have non-Abelian statistics derived from Eq.~(\ref{eq:M2}) using the fact that the non-Abelian quasiparticles carry SU$(2)_2$ charge $1/2$.

\begin{table}[t!]
\[
\begin{array}{|l|c|c|c|c|c|c|}
\hline
\nu = \frac{14}{5}                         &   e^{\ast}   &   \text{n-A?}  &   \theta          &   g_c   &   g_n    &   g     \\
\hline
\overline{\text{L}}_{1/5}\text{:}          &   e/5        &   \text{no}    &   e^{-i \pi /5}   &   1/20  &   1/4    &   3/10  \\
                                           &   3e/5       &   \text{no}    &   e^{i \pi /5}    &   9/20  &   1/4    &   7/10  \\
                                           &   4e/5       &   \text{no}    &   e^{i 4\pi /5}   &   4/5   &   0      &   4/5   \\
\hline
\text{BS}_{4/5}\text{:}                    &   e/5        &   \text{no}    &   e^{i 4\pi /5}    &   1/20  &   3/4    &   4/5   \\
                                           &   e/5        &   \text{yes}   &   e^{i 37\pi /40}  &   1/20  &   7/8    &   37/40 \\
                                           &   2e/5       &   \text{yes}   &   e^{i 13\pi /40}  &   1/5   &   1/8    &   13/40 \\
                                           &   4e/5       &   \text{no}    &   e^{i 4\pi /5}    &   4/5   &   0      &   4/5   \\
\hline
\overline{\text{BS}}_{1/5}^{\psi} \text{:} &   e/5        &   \text{no}    &   e^{i 4\pi /5}    &   1/20  &   3/4    &   4/5   \\
                                           &   e/5        &   \text{yes}   &   e^{i 17\pi /40}  &   1/20  &   5/8    &   27/40 \\
                                           &   2e/5       &   \text{yes}   &   e^{i 7\pi /40}   &   1/5   &   3/8    &   23/40 \\
                                           &   4e/5       &   \text{no}    &   e^{i 4\pi /5}    &   4/5   &   0      &   4/5   \\
\hline
\end{array}
\]
\caption{Relevant quasiparticle excitations of model FQH states at $\nu = 14/5$. Here we list their values of charge $e^{\ast}$; whether they are non-Abelian; their topological twist factor $\theta$; and their charge and neutral scaling exponents $g_c$, $g_n$, and $g$. The BS and $\overline{\text{BS}}_{1/5}^{\psi}$ states are non-Abelian, while the $\overline{\text{L}}$ state is Abelian. All of these have Abelian $4e/5$ Laughlin-type quasiparticles.}
\label{Table_14_5:qps}
\end{table}

We see in Table~\ref{Table_14_5:qps} that all of these states have Abelian $e/5$ and $4e/5$ excitations with relevant scaling exponents, so there should always be a background of such excitations in tunneling. The $e/5$ excitation is the most relevant tunneler for the $\overline{\text{L}}_{1/5}$ state. For the BS$_{4/5}$ and $\overline{\text{BS}}_{1/5}^{\psi}$ states, the non-Abelian $2e/5$ excitation has the most relevant tunneling.

\subsection{$\nu = 19/8$}

While a fully developed FQH plateau has not been observed at $\nu=19/8$, there does appear to be a feature of a developing plateau there~\cite{Xia04,Pan08}, and it is the next filling fraction in the sequence of BS states following $\nu=12/5$, so we will list these states~\cite{Bonderson07d}. These BS states again have non-Abelian statistics derived from Eq.~(\ref{eq:M2}) using the fact that the non-Abelian quasiparticles carry SU$(2)_2$ charge $1/2$.

\begin{table}[t!]
\[
\begin{array}{|l|c|c|c|c|c|c|}
\hline
\nu = \frac{19}{8}                         &   e^{\ast}   &   \text{n-A?}  &   \theta          &   g_c    &   g_n    &   g      \\
\hline
\text{BS}_{3/8}\text{:}                    &   e/16       &   \text{yes}   &   e^{-i 17\pi /32} &   1/96  &   19/24  &   77/96  \\
                                           &   e/8        &   \text{no}    &   e^{-i 5\pi /8}   &   1/24  &   2/3    &   17/24  \\
                                           &   3e/16      &   \text{yes}   &   e^{i 7\pi /32}   &   3/32  &   1/8    &   7/32   \\
                                           &   e/4        &   \text{no}    &   e^{-i \pi /2}    &   1/6   &   2/3    &   5/6    \\
                                           &   3e/8       &   \text{no}    &   e^{i 3\pi /8}    &   3/8   &   0      &   3/8    \\
\hline
\overline{\text{BS}}_{5/8}^{\psi}\text{:}  &   e/8        &   \text{no}    &   e^{-i 5\pi /8}   &   1/24  &   2/3    &   17/24  \\
                                           &   3e/16      &   \text{yes}   &   e^{-i 9\pi /32}  &   3/32  &   3/8    &   15/32  \\
                                           &   e/4        &   \text{no}    &   e^{-i \pi /2}    &   1/6   &   2/3    &   5/6    \\
                                           &   3e/8       &   \text{no}    &   e^{i 3\pi /8}    &   3/8   &   0      &   3/8    \\
\hline
\end{array}
\]
\caption{Relevant quasiparticle excitations of model FQH states at $\nu = 19/8$. Here we list their values of charge $e^{\ast}$; whether they are non-Abelian; their topological twist factor $\theta$; and their charge and neutral scaling exponents $g_c$, $g_n$, and $g$. The BS and $\overline{\text{BS}}^{\psi}$ states are non-Abelian. Both of these have Abelian $3e/8$ Laughlin-type quasiparticles.}
\label{Table_19_8:qps}
\end{table}

We see in Table~\ref{Table_19_8:qps} that for both of these states the two smallest tunneling exponents belong to a non-Abelian $3e/16$ excitation and an Abelian $3e/8$ excitation. The scaling exponent of the Abelian $3e/8$ excitation is $g=3/8$ in both theories, while the non-Abelian $3e/16$ excitation has tunneling exponent $g=7/32$ for the BS state and $g=15/32$ for the $\overline{\text{BS}}^{\psi}$ state. Thus, the contribution of the non-Abelian $3e/16$ excitation will be slightly stronger than that of the Abelian $3e/8$ excitation for the BS state, whereas it will be the other way around for the $\overline{\text{BS}}^{\psi}$ state.

We also mention that a BS type hierarchy could be built over the SU$(2)_2$ NAF state to produce candidates for all the filling fractions listed above~\cite{Bonderson07d}. The relevant data could be read off the above tables for the non-Abelian quasiparticle excitations in the BS states by simply adding $1/4$ to $g_n$ and $g$, and multiplying the twist factors by $e^{i \pi /4}$.

\section{Charge $e/4$ and $e/2$ Backscattering Matrix Elements}
\label{sec:matrix-elements}

In this section, we examine the tunneling amplitudes of $e/4$ and $e/2$ quasiparticles in detail, and argue that generically $\Gamma_{e/4} \gg \Gamma_{e/2}$. When they are small, the tunneling amplitudes
$\Gamma_{e/4}$ and $\Gamma_{e/2}$ are the matrix elements
for the transfer of charge from one edge of a Hall
device to the other. For simplicity and
concreteness, let us suppose that the device is
a Hall bar with a single constriction. Then, the transfer
of charge $q$ from one edge to the other entails a momentum
change $\Delta {k_x} \sim (q/e)\,{\Delta y}/{\ell_0^2}$,
where the $x$-direction is along the Hall bar.
(This is seen most easily in Landau gauge, as we discuss
below in the context of specific trial wavefunctions.)
However, in order to cause a momentum change
of $\Delta {k_x}$, the potential due to the gates
must have weight at this wavevector, i.e. the matrix
element is determined by the variation of the potential
on a length scale $\Delta x \sim 1/\Delta {k_x}\sim (e/q)\,{\ell_0^2}/{\Delta y}$.
Hence, in order to transfer charge $q$ from one edge to the other,
we need the potential to vary on a length scale $\Delta x$ satisfying
\begin{equation}
\Delta x \cdot \Delta y \sim (e/q)\,{\ell_0^2}
\end{equation}
If the constriction were much smaller than this, then
we would expect that the potential would have comparable
weight at the wavevector necessary for charge $e/4$ transfer,
$\Delta {k_x} \sim \,{\Delta y}/4{\ell_0^2}$, and
at the larger wavevector necessary for charge $e/2$ transfer,
$\Delta {k_x} \sim \,{\Delta y}/2{\ell_0^2}$. Otherwise,
we expect the weight to fall off rapidly with the wavevector,
and to have $\Gamma_{e/4}\gg \Gamma_{e/2}$.

This can be made a little more precise by considering,
for the sake of concreteness, the MR Pfaffian state.
We work in Landau gauge on a cylinder \cite{Milovanovic96}:
\begin{equation}
{\Psi_0} = S({Z_1},\ldots,{Z_n}) \,
\text{Pf}\left(\frac{1}{{Z_i} - {Z_j}}\right)
\prod_{i>j} \left({Z_i} - {Z_j}\right)^2 e^{{\sum_i}{y_i^2}/2\ell_0^2}
\end{equation}
where ${Z_i}=e^{i({x_i}+i{y_i})/r}$, $x_i$ and $y_i$
are the coordinates around and along the cylinder,
respectively, and $r$ is the radius of the cylinder.
$S({Z_1},\ldots,{Z_n})$ is a symmetric polynomial
which deforms the shape of the Hall droplet from a
rotationally-symmetric band around the cylinder
to one with a constriction. For instance, we could take
$S({Z_1},\ldots,{Z_n})={\prod_i} \left({Z_i} - {\zeta_1}\right)^p
\left({Z_i} - {\zeta_2}\right)^p$ where ${\zeta_1}$ and
${\zeta_2}$ are points {\it outside} the droplet with the same
$x$-coordinate. The precise form of $S({Z_1},\ldots,{Z_n})$
is not important at the present level of discussion,
but we will assume that it is a polynomial of degree
$q$ which is less than $2N_e$.
Then, the wavefunction
\begin{multline}
\Psi_{1/4} = S({Z_1},\ldots,{Z_n}) \:
\text{Pf}\left(\frac{{Z_i} + {Z_j}}{{Z_i} - {Z_j}}\right)\,\times\\
\prod_{i>j} \left({Z_i} - {Z_j}\right)^2 e^{{\sum_i}{y_i^2}/2\ell_0^2}
\end{multline}
has charge $e/4$ transferred from one edge to the other,
while
\begin{multline}
\Psi_{1/2} = S({Z_1},\ldots,{Z_n}) \:
{\prod_i} {Z_i}\,\times\\
\text{Pf}\left(\frac{1}{{Z_i} - {Z_j}}\right)
\prod_{i>j} \left({Z_i} - {Z_j}\right)^2 e^{{\sum_i}{y_i^2}/2\ell_0^2}
\end{multline}
has charge $e/2$ transferred from one edge to the other.

The tunneling matrix elements $\Gamma_{e/4}$ and $\Gamma_{e/2}$
for charge-$e/4$ and $e/2$ quasiparticles, respectively, are
\begin{equation}
\label{eqn:t-quarter-half}
\Gamma_{e/4} = \bigl\langle \Psi_{1/4} \bigl|  \hat{V}
\bigr|  \Psi_{0}\bigr\rangle \,\, , \hskip 0.4 cm
\Gamma_{e/2} =  \bigl\langle \Psi_{1/2}  \bigl|  \hat{V}
\bigr|  \Psi_{0}\bigr\rangle
\end{equation}
where
\begin{equation}
 \hat{V} = \int dx\,dy\, V(x,y) \, {\sum_i} \delta^{(2)}(z-{z_i})
\end{equation}
and $V(x,y)$ is the potential due to the gates
which define the point contact.

While we would need a detailed knowledge of
$V(x,y)$ and of the precise shape of the Hall droplet
in order to determine $\Gamma_{e/4}$ and $\Gamma_{e/2}$ quantitatively,
we can make a few qualitative remarks which echo our earlier
observations. First, let us ignore $S({Z_1},\ldots,{Z_n})$.
Then,  $\Psi_{0},  \Psi_{1/4}, \Psi_{1/2}$
are eigenstates of angular momentum around the
cylinder with eigenvalues $M={M_0}, {M_0}+N/2,
{M_0}+N$. Thus, the tunneling matrix elements
$\Gamma_{e/4}$ and $\Gamma_{e/2}$ are controlled by
$\tilde{V}({k_x},y)$ for ${k_x}=N/2r$ and
${k_x}=N/r$, respectively. These will be comparable
if the scale $\Delta x$ over which the potential varies
in the $x$-direction is smaller than $1/{k_x} \sim r/N$.
But the distance between the two edges
$\Delta y$ is $\Delta y \sim {\ell_0^2} N/r$.
Hence, we need $\Delta x \cdot\Delta y \sim {\ell_0^2}$
in order for the two tunneling matrix elements to be
comparable. Otherwise, both are determined by the tails
of the (Fourier transform of the) potential and $\Gamma_{e/4}\gg \Gamma_{e/2}$.
The presence of the constrictions, which is reflected
in $S({Z_1},\ldots,{Z_n})$ means that the wavefunctions
are no longer angular momentum eigenstates.
Instead, $\Psi_{0}$ has non-zero amplitude for a range
of angular momenta ${M_0}<M<{M_0}+m$ while
$\Psi_{1/4}$ has non-zero amplitude for a range
${M_0}+N/2<M<{M_0}+N/2+m$, and similarly for
$\Psi_{1/2}$. Here, $m$ is determined by $S({Z_1},\ldots,{Z_n})$;
the minimum distance between the two edges at the constriction is
$\Delta y \sim {\ell_0^2} (N-m)/r$.
Thus, the tunneling matrix elements
$\Gamma_{e/4}$ and $\Gamma_{e/2}$ are controlled by ${k_x}=(N-2m)/2r$
and ${k_x}=(N-m)/r$. Hence, we obtain
the same requirement as above, but with $\Delta y$
now understood as the distance between the two edges
at their point of closest approach.


\begin{thebibliography}{86}
\expandafter\ifx\csname natexlab\endcsname\relax\def\natexlab#1{#1}\fi
\expandafter\ifx\csname bibnamefont\endcsname\relax
  \def\bibnamefont#1{#1}\fi
\expandafter\ifx\csname bibfnamefont\endcsname\relax
  \def\bibfnamefont#1{#1}\fi
\expandafter\ifx\csname citenamefont\endcsname\relax
  \def\citenamefont#1{#1}\fi
\expandafter\ifx\csname url\endcsname\relax
  \def\url#1{\texttt{#1}}\fi
\expandafter\ifx\csname urlprefix\endcsname\relax\def\urlprefix{URL }\fi
\providecommand{\bibinfo}[2]{#2}
\providecommand{\eprint}[2][]{\url{#2}}

\bibitem[{\citenamefont{Venema}(2008)}]{Dreamweaver08}
\bibinfo{author}{\bibfnamefont{L.}~\bibnamefont{Venema}},
  \bibinfo{journal}{Nature News} \textbf{\bibinfo{volume}{452}},
  \bibinfo{pages}{803} (\bibinfo{year}{2008}).

\bibitem[{\citenamefont{Willett et~al.}(1987)\citenamefont{Willett, Eisenstein,
  Stormer, Tsui, Gossard, and English}}]{Willett87}
\bibinfo{author}{\bibfnamefont{R.}~\bibnamefont{Willett}},
  \bibinfo{author}{\bibfnamefont{J.~P.} \bibnamefont{Eisenstein}},
  \bibinfo{author}{\bibfnamefont{H.~L.} \bibnamefont{Stormer}},
  \bibinfo{author}{\bibfnamefont{D.~C.} \bibnamefont{Tsui}},
  \bibinfo{author}{\bibfnamefont{A.~C.} \bibnamefont{Gossard}},
  \bibnamefont{and} \bibinfo{author}{\bibfnamefont{J.~H.}
  \bibnamefont{English}}, \bibinfo{journal}{Phys. Rev. Lett.}
  \textbf{\bibinfo{volume}{59}}, \bibinfo{pages}{1776} (\bibinfo{year}{1987}).

\bibitem[{\citenamefont{Pan et~al.}(1999)\citenamefont{Pan, Xia, Shvarts,
  Adams, Stormer, Tsui, Pfeiffer, Baldwin, and West}}]{Pan99}
\bibinfo{author}{\bibfnamefont{W.}~\bibnamefont{Pan}},
  \bibinfo{author}{\bibfnamefont{J.-S.} \bibnamefont{Xia}},
  \bibinfo{author}{\bibfnamefont{V.}~\bibnamefont{Shvarts}},
  \bibinfo{author}{\bibfnamefont{D.~E.} \bibnamefont{Adams}},
  \bibinfo{author}{\bibfnamefont{H.~L.} \bibnamefont{Stormer}},
  \bibinfo{author}{\bibfnamefont{D.~C.} \bibnamefont{Tsui}},
  \bibinfo{author}{\bibfnamefont{L.~N.} \bibnamefont{Pfeiffer}},
  \bibinfo{author}{\bibfnamefont{K.~W.} \bibnamefont{Baldwin}},
  \bibnamefont{and} \bibinfo{author}{\bibfnamefont{K.~W.} \bibnamefont{West}},
  \bibinfo{journal}{Phys. Rev. Lett.} \textbf{\bibinfo{volume}{83}},
  \bibinfo{pages}{3530} (\bibinfo{year}{1999}), \eprint{cond-mat/9907356}.

\bibitem[{\citenamefont{Greiter et~al.}(1992)\citenamefont{Greiter, Wen, and
  Wilczek}}]{Greiter92}
\bibinfo{author}{\bibfnamefont{M.}~\bibnamefont{Greiter}},
  \bibinfo{author}{\bibfnamefont{X.~G.} \bibnamefont{Wen}}, \bibnamefont{and}
  \bibinfo{author}{\bibfnamefont{F.}~\bibnamefont{Wilczek}},
  \bibinfo{journal}{Nucl. Phys. B} \textbf{\bibinfo{volume}{374}},
  \bibinfo{pages}{567} (\bibinfo{year}{1992}).

\bibitem[{\citenamefont{Moore and Read}(1991)}]{Moore91}
\bibinfo{author}{\bibfnamefont{G.}~\bibnamefont{Moore}} \bibnamefont{and}
  \bibinfo{author}{\bibfnamefont{N.}~\bibnamefont{Read}},
  \bibinfo{journal}{Nucl. Phys. B} \textbf{\bibinfo{volume}{360}},
  \bibinfo{pages}{362} (\bibinfo{year}{1991}).

\bibitem[{\citenamefont{Nayak and Wilczek}(1996)}]{Nayak96c}
\bibinfo{author}{\bibfnamefont{C.}~\bibnamefont{Nayak}} \bibnamefont{and}
  \bibinfo{author}{\bibfnamefont{F.}~\bibnamefont{Wilczek}},
  \bibinfo{journal}{Nucl. Phys. B} \textbf{\bibinfo{volume}{479}},
  \bibinfo{pages}{529} (\bibinfo{year}{1996}), \eprint{cond-mat/9605145}.

\bibitem[{\citenamefont{Read and Rezayi}(1996)}]{Read96}
\bibinfo{author}{\bibfnamefont{N.}~\bibnamefont{Read}} \bibnamefont{and}
  \bibinfo{author}{\bibfnamefont{E.}~\bibnamefont{Rezayi}},
  \bibinfo{journal}{Phys. Rev. B} \textbf{\bibinfo{volume}{54}},
  \bibinfo{pages}{16864} (\bibinfo{year}{1996}), \eprint{cond-mat/9609079}.

\bibitem[{\citenamefont{Leinaas and Myrheim}(1977)}]{Leinaas77}
\bibinfo{author}{\bibfnamefont{J.~M.} \bibnamefont{Leinaas}} \bibnamefont{and}
  \bibinfo{author}{\bibfnamefont{J.}~\bibnamefont{Myrheim}},
  \bibinfo{journal}{Nuovo Cimento B} \textbf{\bibinfo{volume}{37}},
  \bibinfo{pages}{1} (\bibinfo{year}{1977}).

\bibitem[{\citenamefont{Goldin et~al.}(1985)\citenamefont{Goldin, Menikoff, and
  Sharp}}]{Goldin85}
\bibinfo{author}{\bibfnamefont{G.~A.} \bibnamefont{Goldin}},
  \bibinfo{author}{\bibfnamefont{R.}~\bibnamefont{Menikoff}}, \bibnamefont{and}
  \bibinfo{author}{\bibfnamefont{D.~H.} \bibnamefont{Sharp}},
  \bibinfo{journal}{Phys. Rev. Lett.} \textbf{\bibinfo{volume}{54}},
  \bibinfo{pages}{603} (\bibinfo{year}{1985}).

\bibitem[{\citenamefont{Fredenhagen et~al.}(1989)\citenamefont{Fredenhagen,
  Rehren, and Schroer}}]{Fredenhagen89}
\bibinfo{author}{\bibfnamefont{K.}~\bibnamefont{Fredenhagen}},
  \bibinfo{author}{\bibfnamefont{K.~H.} \bibnamefont{Rehren}},
  \bibnamefont{and} \bibinfo{author}{\bibfnamefont{B.}~\bibnamefont{Schroer}},
  \bibinfo{journal}{Commun. Math. Phys.} \textbf{\bibinfo{volume}{125}},
  \bibinfo{pages}{201} (\bibinfo{year}{1989}).

\bibitem[{\citenamefont{Imbo et~al.}(1990)\citenamefont{Imbo, Imbo, and
  Sudarshan}}]{Imbo89}
\bibinfo{author}{\bibfnamefont{T.~D.} \bibnamefont{Imbo}},
  \bibinfo{author}{\bibfnamefont{C.~S.} \bibnamefont{Imbo}}, \bibnamefont{and}
  \bibinfo{author}{\bibfnamefont{E.~C.~G.} \bibnamefont{Sudarshan}},
  \bibinfo{journal}{Phys. Lett.} \textbf{\bibinfo{volume}{B234}},
  \bibinfo{pages}{103} (\bibinfo{year}{1990}).

\bibitem[{\citenamefont{Fr\"{o}hlich and Gabbiani}(1990)}]{Froehlich90}
\bibinfo{author}{\bibfnamefont{J.}~\bibnamefont{Fr\"{o}hlich}}
  \bibnamefont{and} \bibinfo{author}{\bibfnamefont{F.}~\bibnamefont{Gabbiani}},
  \bibinfo{journal}{Rev. Math. Phys.} \textbf{\bibinfo{volume}{2}},
  \bibinfo{pages}{251} (\bibinfo{year}{1990}).

\bibitem[{\citenamefont{Imbo and March-Russell}(1990)}]{Imbo90}
\bibinfo{author}{\bibfnamefont{T.~D.} \bibnamefont{Imbo}} \bibnamefont{and}
  \bibinfo{author}{\bibfnamefont{J.}~\bibnamefont{March-Russell}},
  \bibinfo{journal}{Phys. Lett. B} \textbf{\bibinfo{volume}{252}},
  \bibinfo{pages}{84} (\bibinfo{year}{1990}).

\bibitem[{\citenamefont{Bais et~al.}(1992)\citenamefont{Bais, van Driel, and
  de~Wild~Propitius}}]{Bais92}
\bibinfo{author}{\bibfnamefont{F.~A.} \bibnamefont{Bais}},
  \bibinfo{author}{\bibfnamefont{P.}~\bibnamefont{van Driel}},
  \bibnamefont{and}
  \bibinfo{author}{\bibfnamefont{M.}~\bibnamefont{de~Wild~Propitius}},
  \bibinfo{journal}{Phys. Lett. B} \textbf{\bibinfo{volume}{280}},
  \bibinfo{pages}{63} (\bibinfo{year}{1992}), \eprint{hep-th/9203046}.

\bibitem[{\citenamefont{Kitaev}(2003)}]{Kitaev03}
\bibinfo{author}{\bibfnamefont{A.~Y.} \bibnamefont{Kitaev}},
  \bibinfo{journal}{Ann. Phys.} \textbf{\bibinfo{volume}{303}},
  \bibinfo{pages}{2} (\bibinfo{year}{2003}), \eprint{quant-ph/9707021}.

\bibitem[{\citenamefont{Preskill}(1998)}]{Preskill98}
\bibinfo{author}{\bibfnamefont{J.}~\bibnamefont{Preskill}}, in
  \emph{\bibinfo{booktitle}{Introduction to Quantum Computation}}, edited by
  \bibinfo{editor}{\bibfnamefont{H.-K.} \bibnamefont{Lo}},
  \bibinfo{editor}{\bibfnamefont{S.}~\bibnamefont{Popescu}}, \bibnamefont{and}
  \bibinfo{editor}{\bibfnamefont{T.~P.} \bibnamefont{Spiller}}
  (\bibinfo{publisher}{World Scientific}, \bibinfo{year}{1998}),
  \eprint{quant-ph/9712048}.

\bibitem[{\citenamefont{Freedman}(1998)}]{Freedman98}
\bibinfo{author}{\bibfnamefont{M.~H.} \bibnamefont{Freedman}},
  \bibinfo{journal}{Proc. Natl. Acad. Sci. USA} \textbf{\bibinfo{volume}{95}},
  \bibinfo{pages}{98} (\bibinfo{year}{1998}).

\bibitem[{\citenamefont{Freedman
  et~al.}(2002{\natexlab{a}})\citenamefont{Freedman, Larsen, and
  Wang}}]{Freedman02a}
\bibinfo{author}{\bibfnamefont{M.~H.} \bibnamefont{Freedman}},
  \bibinfo{author}{\bibfnamefont{M.~J.} \bibnamefont{Larsen}},
  \bibnamefont{and} \bibinfo{author}{\bibfnamefont{Z.}~\bibnamefont{Wang}},
  \bibinfo{journal}{Commun. Math. Phys.} \textbf{\bibinfo{volume}{227}},
  \bibinfo{pages}{605} (\bibinfo{year}{2002}{\natexlab{a}}),
  \eprint{quant-ph/0001108}.

\bibitem[{\citenamefont{Freedman
  et~al.}(2002{\natexlab{b}})\citenamefont{Freedman, Larsen, and
  Wang}}]{Freedman02b}
\bibinfo{author}{\bibfnamefont{M.~H.} \bibnamefont{Freedman}},
  \bibinfo{author}{\bibfnamefont{M.~J.} \bibnamefont{Larsen}},
  \bibnamefont{and} \bibinfo{author}{\bibfnamefont{Z.}~\bibnamefont{Wang}},
  \bibinfo{journal}{Commun. Math. Phys.} \textbf{\bibinfo{volume}{228}},
  \bibinfo{pages}{177} (\bibinfo{year}{2002}{\natexlab{b}}),
  \eprint{math/0103200}.

\bibitem[{\citenamefont{Freedman et~al.}(2003)\citenamefont{Freedman, Kitaev,
  Larsen, and Wang}}]{Freedman03b}
\bibinfo{author}{\bibfnamefont{M.~H.} \bibnamefont{Freedman}},
  \bibinfo{author}{\bibfnamefont{A.}~\bibnamefont{Kitaev}},
  \bibinfo{author}{\bibfnamefont{M.~J.} \bibnamefont{Larsen}},
  \bibnamefont{and} \bibinfo{author}{\bibfnamefont{Z.}~\bibnamefont{Wang}},
  \bibinfo{journal}{Bull. Amer. Math. Soc. (N.S.)}
  \textbf{\bibinfo{volume}{40}}, \bibinfo{pages}{31} (\bibinfo{year}{2002}),
  \eprint{quant-ph/0101025}.

\bibitem[{\citenamefont{Preskill}(2004)}]{Preskill-lectures}
\bibinfo{author}{\bibfnamefont{J.}~\bibnamefont{Preskill}}
  (\bibinfo{year}{2004}), \bibinfo{note}{lecture notes},
  \urlprefix\url{http://www.theory.caltech.edu/~preskill/ph219/topological.ps}.

\bibitem[{\citenamefont{Nayak et~al.}(2008)\citenamefont{Nayak, Simon, Stern,
  Freedman, and Das~Sarma}}]{Nayak08}
\bibinfo{author}{\bibfnamefont{C.}~\bibnamefont{Nayak}},
  \bibinfo{author}{\bibfnamefont{S.~H.} \bibnamefont{Simon}},
  \bibinfo{author}{\bibfnamefont{A.}~\bibnamefont{Stern}},
  \bibinfo{author}{\bibfnamefont{M.}~\bibnamefont{Freedman}}, \bibnamefont{and}
  \bibinfo{author}{\bibfnamefont{S.}~\bibnamefont{Das~Sarma}},
  \bibinfo{journal}{Rev. Mod. Phys.} \textbf{\bibinfo{volume}{80}},
  \bibinfo{pages}{1083} (\bibinfo{year}{2008}), \eprint{arXiv:0707.1889}.

\bibitem[{\citenamefont{Dolev et~al.}(2008)\citenamefont{Dolev, Heiblum,
  Umansky, Stern, and Mahalu}}]{Dolev08}
\bibinfo{author}{\bibfnamefont{M.}~\bibnamefont{Dolev}},
  \bibinfo{author}{\bibfnamefont{M.}~\bibnamefont{Heiblum}},
  \bibinfo{author}{\bibfnamefont{V.}~\bibnamefont{Umansky}},
  \bibinfo{author}{\bibfnamefont{A.}~\bibnamefont{Stern}}, \bibnamefont{and}
  \bibinfo{author}{\bibfnamefont{D.}~\bibnamefont{Mahalu}},
  \bibinfo{journal}{Nature} \textbf{\bibinfo{volume}{452}},
  \bibinfo{pages}{829} (\bibinfo{year}{2008}), \eprint{ar{X}iv:0802.0930}.

\bibitem[{\citenamefont{Radu et~al.}(2008)\citenamefont{Radu, Miller, Marcus,
  Kastner, Pfeiffer, and West}}]{Radu08}
\bibinfo{author}{\bibfnamefont{I.~P.} \bibnamefont{Radu}},
  \bibinfo{author}{\bibfnamefont{J.~B.} \bibnamefont{Miller}},
  \bibinfo{author}{\bibfnamefont{C.~M.} \bibnamefont{Marcus}},
  \bibinfo{author}{\bibfnamefont{M.~A.} \bibnamefont{Kastner}},
  \bibinfo{author}{\bibfnamefont{L.~N.} \bibnamefont{Pfeiffer}},
  \bibnamefont{and} \bibinfo{author}{\bibfnamefont{K.~W.} \bibnamefont{West}},
  \bibinfo{journal}{Science} \textbf{\bibinfo{volume}{320}},
  \bibinfo{pages}{899} (\bibinfo{year}{2008}), \eprint{ar{X}iv:0803.3530}.

\bibitem[{\citenamefont{Lee et~al.}(2007)\citenamefont{Lee, Ryu, Nayak, and
  Fisher}}]{Lee07}
\bibinfo{author}{\bibfnamefont{S.-S.} \bibnamefont{Lee}},
  \bibinfo{author}{\bibfnamefont{S.}~\bibnamefont{Ryu}},
  \bibinfo{author}{\bibfnamefont{C.}~\bibnamefont{Nayak}}, \bibnamefont{and}
  \bibinfo{author}{\bibfnamefont{M.~P.~A.} \bibnamefont{Fisher}},
  \bibinfo{journal}{Phys. Rev. Lett.} \textbf{\bibinfo{volume}{99}},
  \bibinfo{pages}{236807} (\bibinfo{year}{2007}), \eprint{arXiv:0707.0478}.

\bibitem[{\citenamefont{Levin et~al.}(2007)\citenamefont{Levin, Halperin, and
  Rosenow}}]{Levin07}
\bibinfo{author}{\bibfnamefont{M.}~\bibnamefont{Levin}},
  \bibinfo{author}{\bibfnamefont{B.~I.} \bibnamefont{Halperin}},
  \bibnamefont{and} \bibinfo{author}{\bibfnamefont{B.}~\bibnamefont{Rosenow}},
  \bibinfo{journal}{Phys. Rev. Lett.} \textbf{\bibinfo{volume}{99}},
  \bibinfo{pages}{236806} (\bibinfo{year}{2007}), \eprint{arXiv:0707.0483}.

\bibitem[{\citenamefont{Wen}(1991)}]{Wen91a}
\bibinfo{author}{\bibfnamefont{X.~G.} \bibnamefont{Wen}},
  \bibinfo{journal}{Phys. Rev. Lett.} \textbf{\bibinfo{volume}{66}},
  \bibinfo{pages}{802} (\bibinfo{year}{1991}).

\bibitem[{\citenamefont{Blok and Wen}(1992)}]{Blok92}
\bibinfo{author}{\bibfnamefont{B.}~\bibnamefont{Blok}} \bibnamefont{and}
  \bibinfo{author}{\bibfnamefont{X.~G.} \bibnamefont{Wen}},
  \bibinfo{journal}{Nucl. Phys. B} \textbf{\bibinfo{volume}{374}},
  \bibinfo{pages}{615} (\bibinfo{year}{1992}).

\bibitem[{\citenamefont{Halperin}(1983)}]{Halperin83}
\bibinfo{author}{\bibfnamefont{B.~I.} \bibnamefont{Halperin}},
  \bibinfo{journal}{Helv. Phys. Acta} \textbf{\bibinfo{volume}{56}},
  \bibinfo{pages}{75} (\bibinfo{year}{1983}).

\bibitem[{\citenamefont{de~C.~Chamon et~al.}(1997)\citenamefont{de~C.~Chamon,
  Freed, Kivelson, Sondhi, and Wen}}]{Chamon97}
\bibinfo{author}{\bibfnamefont{C.}~\bibnamefont{de~C.~Chamon}},
  \bibinfo{author}{\bibfnamefont{D.~E.} \bibnamefont{Freed}},
  \bibinfo{author}{\bibfnamefont{S.~A.} \bibnamefont{Kivelson}},
  \bibinfo{author}{\bibfnamefont{S.~L.} \bibnamefont{Sondhi}},
  \bibnamefont{and} \bibinfo{author}{\bibfnamefont{X.~G.} \bibnamefont{Wen}},
  \bibinfo{journal}{Phys. Rev. B} \textbf{\bibinfo{volume}{55}},
  \bibinfo{pages}{2331} (\bibinfo{year}{1997}), \eprint{cond-mat/9607195}.

\bibitem[{\citenamefont{Fradkin et~al.}(1998)\citenamefont{Fradkin, Nayak,
  Tsvelik, and Wilczek}}]{Fradkin98}
\bibinfo{author}{\bibfnamefont{E.}~\bibnamefont{Fradkin}},
  \bibinfo{author}{\bibfnamefont{C.}~\bibnamefont{Nayak}},
  \bibinfo{author}{\bibfnamefont{A.}~\bibnamefont{Tsvelik}}, \bibnamefont{and}
  \bibinfo{author}{\bibfnamefont{F.}~\bibnamefont{Wilczek}},
  \bibinfo{journal}{Nucl. Phys. B} \textbf{\bibinfo{volume}{516}},
  \bibinfo{pages}{704} (\bibinfo{year}{1998}), \eprint{cond-mat/9711087}.

\bibitem[{\citenamefont{Das~Sarma et~al.}(2005)\citenamefont{Das~Sarma,
  Freedman, and Nayak}}]{DasSarma05}
\bibinfo{author}{\bibfnamefont{S.}~\bibnamefont{Das~Sarma}},
  \bibinfo{author}{\bibfnamefont{M.}~\bibnamefont{Freedman}}, \bibnamefont{and}
  \bibinfo{author}{\bibfnamefont{C.}~\bibnamefont{Nayak}},
  \bibinfo{journal}{Phys. Rev. Lett.} \textbf{\bibinfo{volume}{94}},
  \bibinfo{pages}{166802} (\bibinfo{year}{2005}), \eprint{cond-mat/0412343}.

\bibitem[{\citenamefont{Stern and Halperin}(2006)}]{Stern06a}
\bibinfo{author}{\bibfnamefont{A.}~\bibnamefont{Stern}} \bibnamefont{and}
  \bibinfo{author}{\bibfnamefont{B.~I.} \bibnamefont{Halperin}},
  \bibinfo{journal}{Phys. Rev. Lett.} \textbf{\bibinfo{volume}{96}},
  \bibinfo{pages}{016802} (\bibinfo{year}{2006}), \eprint{cond-mat/0508447}.

\bibitem[{\citenamefont{Bonderson
  et~al.}(2006{\natexlab{a}})\citenamefont{Bonderson, Kitaev, and
  Shtengel}}]{Bonderson06a}
\bibinfo{author}{\bibfnamefont{P.}~\bibnamefont{Bonderson}},
  \bibinfo{author}{\bibfnamefont{A.}~\bibnamefont{Kitaev}}, \bibnamefont{and}
  \bibinfo{author}{\bibfnamefont{K.}~\bibnamefont{Shtengel}},
  \bibinfo{journal}{Phys. Rev. Lett.} \textbf{\bibinfo{volume}{96}},
  \bibinfo{pages}{016803} (\bibinfo{year}{2006}{\natexlab{a}}),
  \eprint{cond-mat/0508616}.

\bibitem[{\citenamefont{Bonderson
  et~al.}(2006{\natexlab{b}})\citenamefont{Bonderson, Shtengel, and
  Slingerland}}]{Bonderson06b}
\bibinfo{author}{\bibfnamefont{P.}~\bibnamefont{Bonderson}},
  \bibinfo{author}{\bibfnamefont{K.}~\bibnamefont{Shtengel}}, \bibnamefont{and}
  \bibinfo{author}{\bibfnamefont{J.~K.} \bibnamefont{Slingerland}},
  \bibinfo{journal}{Phys. Rev. Lett.} \textbf{\bibinfo{volume}{97}},
  \bibinfo{pages}{016401} (\bibinfo{year}{2006}{\natexlab{b}}),
  \eprint{cond-mat/0601242}.

\bibitem[{\citenamefont{Bonderson
  et~al.}(2008{\natexlab{a}})\citenamefont{Bonderson, Freedman, and
  Nayak}}]{Bonderson08a}
\bibinfo{author}{\bibfnamefont{P.}~\bibnamefont{Bonderson}},
  \bibinfo{author}{\bibfnamefont{M.}~\bibnamefont{Freedman}}, \bibnamefont{and}
  \bibinfo{author}{\bibfnamefont{C.}~\bibnamefont{Nayak}},
  \bibinfo{journal}{Phys. Rev. Lett.} \textbf{\bibinfo{volume}{101}},
  \bibinfo{pages}{010501} (\bibinfo{year}{2008}{\natexlab{a}}),
  \eprint{arXiv:0802.0279}.

\bibitem[{\citenamefont{Bonderson
  et~al.}(2009{\natexlab{a}})\citenamefont{Bonderson, Freedman, and
  Nayak}}]{Bonderson08b}
\bibinfo{author}{\bibfnamefont{P.}~\bibnamefont{Bonderson}},
  \bibinfo{author}{\bibfnamefont{M.}~\bibnamefont{Freedman}}, \bibnamefont{and}
  \bibinfo{author}{\bibfnamefont{C.}~\bibnamefont{Nayak}},
  \bibinfo{journal}{Annals of Physics} \textbf{\bibinfo{volume}{324}},
  \bibinfo{pages}{787} (\bibinfo{year}{2009}{\natexlab{a}}),
  \eprint{arXiv:0808.1933}.

\bibitem[{\citenamefont{Ji et~al.}(2003)\citenamefont{Ji, Chung, Sprinzak,
  Heiblum, Mahalu, and Shtrikman}}]{Ji03}
\bibinfo{author}{\bibfnamefont{Y.}~\bibnamefont{Ji}},
  \bibinfo{author}{\bibfnamefont{Y.}~\bibnamefont{Chung}},
  \bibinfo{author}{\bibfnamefont{D.}~\bibnamefont{Sprinzak}},
  \bibinfo{author}{\bibfnamefont{M.}~\bibnamefont{Heiblum}},
  \bibinfo{author}{\bibfnamefont{D.}~\bibnamefont{Mahalu}}, \bibnamefont{and}
  \bibinfo{author}{\bibfnamefont{H.}~\bibnamefont{Shtrikman}},
  \bibinfo{journal}{Nature} \textbf{\bibinfo{volume}{422}},
  \bibinfo{pages}{415} (\bibinfo{year}{2003}), \eprint{cond-mat/0303553}.

\bibitem[{\citenamefont{Zhang et~al.}(2009)\citenamefont{Zhang, McClure,
  Levenson-Falk, Marcus, Pfeiffer, and West}}]{Zhang09}
\bibinfo{author}{\bibfnamefont{Y.}~\bibnamefont{Zhang}},
  \bibinfo{author}{\bibfnamefont{D.~T.} \bibnamefont{McClure}},
  \bibinfo{author}{\bibfnamefont{E.~M.} \bibnamefont{Levenson-Falk}},
  \bibinfo{author}{\bibfnamefont{C.~M.} \bibnamefont{Marcus}},
  \bibinfo{author}{\bibfnamefont{L.~N.} \bibnamefont{Pfeiffer}},
  \bibnamefont{and} \bibinfo{author}{\bibfnamefont{K.~W.} \bibnamefont{West}},
  \bibinfo{journal}{Phys. Rev. B} \textbf{\bibinfo{volume}{79}},
  \bibinfo{pages}{241304(R)} (\bibinfo{year}{2009}), \eprint{arXiv:0901.0127}.

\bibitem[{\citenamefont{Camino et~al.}(2005)\citenamefont{Camino, Zhou, and
  Goldman}}]{Camino05a}
\bibinfo{author}{\bibfnamefont{F.~E.} \bibnamefont{Camino}},
  \bibinfo{author}{\bibfnamefont{W.}~\bibnamefont{Zhou}}, \bibnamefont{and}
  \bibinfo{author}{\bibfnamefont{V.~J.} \bibnamefont{Goldman}},
  \bibinfo{journal}{Phys. Rev. B} \textbf{\bibinfo{volume}{72}},
  \bibinfo{pages}{075342} (\bibinfo{year}{2005}), \eprint{cond-mat/0502406}.

\bibitem[{\citenamefont{Camino et~al.}(2007)\citenamefont{Camino, Zhou, and
  Goldman}}]{Camino07a}
\bibinfo{author}{\bibfnamefont{F.~E.} \bibnamefont{Camino}},
  \bibinfo{author}{\bibfnamefont{W.}~\bibnamefont{Zhou}}, \bibnamefont{and}
  \bibinfo{author}{\bibfnamefont{V.~J.} \bibnamefont{Goldman}},
  \bibinfo{journal}{Phys. Rev. Lett.} \textbf{\bibinfo{volume}{98}},
  \bibinfo{pages}{076805} (\bibinfo{year}{2007}), \eprint{cond-mat/0610751}.

\bibitem[{\citenamefont{Godfrey et~al.}(2007)\citenamefont{Godfrey, Jiang,
  Kang, Simon, Baldwin, Pfeiffer, and West}}]{Godfrey07}
\bibinfo{author}{\bibfnamefont{M.~D.} \bibnamefont{Godfrey}},
  \bibinfo{author}{\bibfnamefont{P.}~\bibnamefont{Jiang}},
  \bibinfo{author}{\bibfnamefont{W.}~\bibnamefont{Kang}},
  \bibinfo{author}{\bibfnamefont{S.~H.} \bibnamefont{Simon}},
  \bibinfo{author}{\bibfnamefont{K.~W.} \bibnamefont{Baldwin}},
  \bibinfo{author}{\bibfnamefont{L.~N.} \bibnamefont{Pfeiffer}},
  \bibnamefont{and} \bibinfo{author}{\bibfnamefont{K.~W.} \bibnamefont{West}}
  (\bibinfo{year}{2007}), \eprint{arxiv:0708.2448}.

\bibitem[{\citenamefont{Willett et~al.}(2008)\citenamefont{Willett, Manfra,
  Pfeiffer, and West}}]{Willett08}
\bibinfo{author}{\bibfnamefont{R.~L.} \bibnamefont{Willett}},
  \bibinfo{author}{\bibfnamefont{M.~J.} \bibnamefont{Manfra}},
  \bibinfo{author}{\bibfnamefont{L.~N.} \bibnamefont{Pfeiffer}},
  \bibnamefont{and} \bibinfo{author}{\bibfnamefont{K.~W.} \bibnamefont{West}}
  (\bibinfo{year}{2008}), \eprint{arXiv:0807.0221v1}.

\bibitem[{\citenamefont{Willett
  et~al.}(2009{\natexlab{a}})\citenamefont{Willett, Pfeiffer, and
  West}}]{Willett09a}
\bibinfo{author}{\bibfnamefont{R.~L.} \bibnamefont{Willett}},
  \bibinfo{author}{\bibfnamefont{L.~N.} \bibnamefont{Pfeiffer}},
  \bibnamefont{and} \bibinfo{author}{\bibfnamefont{K.~W.} \bibnamefont{West}},
  \bibinfo{journal}{Proc. Natl. Acad. Sci.} \textbf{\bibinfo{volume}{106}},
  \bibinfo{pages}{8853} (\bibinfo{year}{2009}{\natexlab{a}}),
  \eprint{arXiv:0807.0221v3}.

\bibitem[{\citenamefont{Willett
  et~al.}(2009{\natexlab{b}})\citenamefont{Willett, Pfeiffer, and
  West}}]{Willett09u}
\bibinfo{author}{\bibfnamefont{R.~L.} \bibnamefont{Willett}},
  \bibinfo{author}{\bibfnamefont{L.~N.} \bibnamefont{Pfeiffer}},
  \bibnamefont{and} \bibinfo{author}{\bibfnamefont{K.~W.} \bibnamefont{West}}
  (\bibinfo{year}{2009}{\natexlab{b}}), \eprint{unpublished}.

\bibitem[{\citenamefont{Bonderson et~al.}(2007)\citenamefont{Bonderson,
  Shtengel, and Slingerland}}]{Bonderson07a}
\bibinfo{author}{\bibfnamefont{P.}~\bibnamefont{Bonderson}},
  \bibinfo{author}{\bibfnamefont{K.}~\bibnamefont{Shtengel}}, \bibnamefont{and}
  \bibinfo{author}{\bibfnamefont{J.~K.} \bibnamefont{Slingerland}},
  \bibinfo{journal}{Phys. Rev. Lett.} \textbf{\bibinfo{volume}{98}},
  \bibinfo{pages}{070401} (\bibinfo{year}{2007}), \eprint{quant-ph/0608119}.

\bibitem[{\citenamefont{Grosfeld et~al.}(2006)\citenamefont{Grosfeld, Simon,
  and Stern}}]{Grosfeld06b}
\bibinfo{author}{\bibfnamefont{E.}~\bibnamefont{Grosfeld}},
  \bibinfo{author}{\bibfnamefont{S.~H.} \bibnamefont{Simon}}, \bibnamefont{and}
  \bibinfo{author}{\bibfnamefont{A.}~\bibnamefont{Stern}},
  \bibinfo{journal}{Phys. Rev. Lett.} \textbf{\bibinfo{volume}{96}},
  \bibinfo{pages}{226803} (\bibinfo{year}{2006}), \eprint{cond-mat/0602634}.

\bibitem[{\citenamefont{Bonderson}(2007)}]{Bonderson07b}
\bibinfo{author}{\bibfnamefont{P.~H.} \bibnamefont{Bonderson}}, Ph.D. thesis, Caltech
  (\bibinfo{year}{2007}).

\bibitem[{\citenamefont{Bonderson
  et~al.}(2008{\natexlab{b}})\citenamefont{Bonderson, Shtengel, and
  Slingerland}}]{Bonderson07c}
\bibinfo{author}{\bibfnamefont{P.}~\bibnamefont{Bonderson}},
  \bibinfo{author}{\bibfnamefont{K.}~\bibnamefont{Shtengel}}, \bibnamefont{and}
  \bibinfo{author}{\bibfnamefont{J.~K.} \bibnamefont{Slingerland}},
  \bibinfo{journal}{Annals of Physics} \textbf{\bibinfo{volume}{323}},
  \bibinfo{pages}{2709} (\bibinfo{year}{2008}{\natexlab{b}}),
  \eprint{arXiv:0707.4206}.

\bibitem[{\citenamefont{Fendley et~al.}(2007)\citenamefont{Fendley, Fisher, and
  Nayak}}]{Fendley07a}
\bibinfo{author}{\bibfnamefont{P.}~\bibnamefont{Fendley}},
  \bibinfo{author}{\bibfnamefont{M.~P.~A.} \bibnamefont{Fisher}},
  \bibnamefont{and} \bibinfo{author}{\bibfnamefont{C.}~\bibnamefont{Nayak}},
  \bibinfo{journal}{Phys. Rev. B} \textbf{\bibinfo{volume}{75}},
  \bibinfo{pages}{045317} (\bibinfo{year}{2007}), \eprint{cond-mat/0607431}.

\bibitem[{\citenamefont{Bishara and Nayak}(2009)}]{Bishara09b}
\bibinfo{author}{\bibfnamefont{W.}~\bibnamefont{Bishara}} \bibnamefont{and}
  \bibinfo{author}{\bibfnamefont{C.}~\bibnamefont{Nayak}}
  (\bibinfo{year}{2009}), \eprint{arXiv:0906.0327}.

\bibitem[{\citenamefont{Rosenow et~al.}(2009)\citenamefont{Rosenow, Halperin,
  Simon, and Stern}}]{Rosenow09}
\bibinfo{author}{\bibfnamefont{B.}~\bibnamefont{Rosenow}},
  \bibinfo{author}{\bibfnamefont{B.~I.} \bibnamefont{Halperin}},
  \bibinfo{author}{\bibfnamefont{S.~H.} \bibnamefont{Simon}}, \bibnamefont{and}
  \bibinfo{author}{\bibfnamefont{A.}~\bibnamefont{Stern}}
  (\bibinfo{year}{2009}), \eprint{arXiv:0906.0310}.

\bibitem[{\citenamefont{Bishara and Nayak}(2008)}]{Bishara08}
\bibinfo{author}{\bibfnamefont{W.}~\bibnamefont{Bishara}} \bibnamefont{and}
  \bibinfo{author}{\bibfnamefont{C.}~\bibnamefont{Nayak}},
  \bibinfo{journal}{Phys. Rev. B} \textbf{\bibinfo{volume}{77}},
  \bibinfo{pages}{165302} (\bibinfo{year}{2008}), \eprint{arXiv:0708.2704}.

\bibitem[{\citenamefont{Fidkowski}(2007)}]{Fidkowski07c}
\bibinfo{author}{\bibfnamefont{L.}~\bibnamefont{Fidkowski}}
  (\bibinfo{year}{2007}), \eprint{arXiv:0704.3291}.

\bibitem[{\citenamefont{Ardonne and Kim}(2008)}]{Ardonne07a}
\bibinfo{author}{\bibfnamefont{E.}~\bibnamefont{Ardonne}} \bibnamefont{and}
  \bibinfo{author}{\bibfnamefont{E.-A.} \bibnamefont{Kim}},
  \bibinfo{journal}{J. Stat. Mech.} p. \bibinfo{pages}{L04001}
  (\bibinfo{year}{2008}), \eprint{arXiv:0705.2902}.

\bibitem[{\citenamefont{Wan et~al.}(2008)\citenamefont{Wan, Hu, Rezayi, and
  Yang}}]{Wan08a}
\bibinfo{author}{\bibfnamefont{X.}~\bibnamefont{Wan}},
  \bibinfo{author}{\bibfnamefont{Z.-X.} \bibnamefont{Hu}},
  \bibinfo{author}{\bibfnamefont{E.~H.} \bibnamefont{Rezayi}},
  \bibnamefont{and} \bibinfo{author}{\bibfnamefont{K.}~\bibnamefont{Yang}},
  \bibinfo{journal}{Phys. Rev. B} \textbf{\bibinfo{volume}{77}},
  \bibinfo{pages}{165316} (\bibinfo{year}{2008}), \eprint{ar{X}iv:0712.2095}.

\bibitem[{\citenamefont{Ilan et~al.}(2008)\citenamefont{Ilan, Grosfeld, and
  Stern}}]{Ilan08a}
\bibinfo{author}{\bibfnamefont{R.}~\bibnamefont{Ilan}},
  \bibinfo{author}{\bibfnamefont{E.}~\bibnamefont{Grosfeld}}, \bibnamefont{and}
  \bibinfo{author}{\bibfnamefont{A.}~\bibnamefont{Stern}},
  \bibinfo{journal}{Phys. Rev. Lett.} \textbf{\bibinfo{volume}{100}},
  \bibinfo{pages}{086803} (\bibinfo{year}{2008}), \eprint{ar{X}iv:0705.2187}.

\bibitem[{\citenamefont{Bonderson
  et~al.}(2009{\natexlab{b}})\citenamefont{Bonderson, Nayak, and
  Shtengel}}]{Bonderson09d}
\bibinfo{author}{\bibfnamefont{P.}~\bibnamefont{Bonderson}},
  \bibinfo{author}{\bibfnamefont{C.}~\bibnamefont{Nayak}}, \bibnamefont{and}
  \bibinfo{author}{\bibfnamefont{K.}~\bibnamefont{Shtengel}}
  (\bibinfo{year}{2009}{\natexlab{b}}), \eprint{ar{X}iv:0909.1056}.

\bibitem[{\citenamefont{Folk et~al.}(1996)\citenamefont{Folk, Patel, Godijn,
  Huibers, Cronenwett, Marcus, Campman, and Gossard}}]{Folk96}
\bibinfo{author}{\bibfnamefont{J.~A.} \bibnamefont{Folk}},
  \bibinfo{author}{\bibfnamefont{S.~R.} \bibnamefont{Patel}},
  \bibinfo{author}{\bibfnamefont{S.~F.} \bibnamefont{Godijn}},
  \bibinfo{author}{\bibfnamefont{A.~G.} \bibnamefont{Huibers}},
  \bibinfo{author}{\bibfnamefont{S.~M.} \bibnamefont{Cronenwett}},
  \bibinfo{author}{\bibfnamefont{C.~M.} \bibnamefont{Marcus}},
  \bibinfo{author}{\bibfnamefont{K.}~\bibnamefont{Campman}}, \bibnamefont{and}
  \bibinfo{author}{\bibfnamefont{A.~C.} \bibnamefont{Gossard}},
  \bibinfo{journal}{Phys. Rev. Lett.} \textbf{\bibinfo{volume}{76}},
  \bibinfo{pages}{1699} (\bibinfo{year}{1996}).

\bibitem[{\citenamefont{Ofek et~al.}()\citenamefont{Ofek, Heiblum, Umansky,
  Stern, and Mahalu}}]{Ofek09}
\bibinfo{author}{\bibfnamefont{N.}~\bibnamefont{Ofek}},
  \bibinfo{author}{\bibfnamefont{M.}~\bibnamefont{Heiblum}},
  \bibinfo{author}{\bibfnamefont{V.}~\bibnamefont{Umansky}},
  \bibinfo{author}{\bibfnamefont{A.}~\bibnamefont{Stern}}, \bibnamefont{and}
  \bibinfo{author}{\bibfnamefont{D.}~\bibnamefont{Mahalu}}.

\bibitem[{\citenamefont{Rosenow and Halperin}(2007)}]{Rosenow07}
\bibinfo{author}{\bibfnamefont{B.}~\bibnamefont{Rosenow}} \bibnamefont{and}
  \bibinfo{author}{\bibfnamefont{B.~I.} \bibnamefont{Halperin}},
  \bibinfo{journal}{Phys. Rev. Lett.} \textbf{\bibinfo{volume}{98}},
  \bibinfo{pages}{106801} (\bibinfo{year}{2007}), \eprint{cond-mat/0611101}.

\bibitem[{\citenamefont{Fendley et~al.}(2006)\citenamefont{Fendley, Fisher, and
  Nayak}}]{Fendley06a}
\bibinfo{author}{\bibfnamefont{P.}~\bibnamefont{Fendley}},
  \bibinfo{author}{\bibfnamefont{M.~P.~A.} \bibnamefont{Fisher}},
  \bibnamefont{and} \bibinfo{author}{\bibfnamefont{C.}~\bibnamefont{Nayak}},
  \bibinfo{journal}{Phys. Rev. Lett.} \textbf{\bibinfo{volume}{97}},
  \bibinfo{pages}{036801} (\bibinfo{year}{2006}), \eprint{cond-mat/0604064}.

\bibitem[{\citenamefont{Wen}(1992)}]{Wen92b}
\bibinfo{author}{\bibfnamefont{X.~G.} \bibnamefont{Wen}},
  \bibinfo{journal}{Intl. J. Mod. Phys. B} \textbf{\bibinfo{volume}{6}},
  \bibinfo{pages}{1711} (\bibinfo{year}{1992}).

\bibitem[{\citenamefont{Martin et~al.}(2004)\citenamefont{Martin, Ilani,
  Verdene, Smet, Umansky, Mahalu, Schuh, Abstreiter, and Yacoby}}]{Martin04}
\bibinfo{author}{\bibfnamefont{J.}~\bibnamefont{Martin}},
  \bibinfo{author}{\bibfnamefont{S.}~\bibnamefont{Ilani}},
  \bibinfo{author}{\bibfnamefont{B.}~\bibnamefont{Verdene}},
  \bibinfo{author}{\bibfnamefont{J.}~\bibnamefont{Smet}},
  \bibinfo{author}{\bibfnamefont{V.}~\bibnamefont{Umansky}},
  \bibinfo{author}{\bibfnamefont{D.}~\bibnamefont{Mahalu}},
  \bibinfo{author}{\bibfnamefont{D.}~\bibnamefont{Schuh}},
  \bibinfo{author}{\bibfnamefont{G.}~\bibnamefont{Abstreiter}},
  \bibnamefont{and} \bibinfo{author}{\bibfnamefont{A.}~\bibnamefont{Yacoby}},
  \bibinfo{journal}{Science} \textbf{\bibinfo{volume}{305}},
  \bibinfo{pages}{980} (\bibinfo{year}{2004}).

\bibitem[{\citenamefont{Morf}(1998)}]{Morf98}
\bibinfo{author}{\bibfnamefont{R.~H.} \bibnamefont{Morf}},
  \bibinfo{journal}{Phys. Rev. Lett.} \textbf{\bibinfo{volume}{80}},
  \bibinfo{pages}{1505} (\bibinfo{year}{1998}), \eprint{cond-mat/9809024}.

\bibitem[{\citenamefont{Rezayi and Haldane}(2000)}]{Rezayi00}
\bibinfo{author}{\bibfnamefont{E.~H.} \bibnamefont{Rezayi}} \bibnamefont{and}
  \bibinfo{author}{\bibfnamefont{F.~D.~M.} \bibnamefont{Haldane}},
  \bibinfo{journal}{Phys. Rev. Lett.} \textbf{\bibinfo{volume}{84}},
  \bibinfo{pages}{4685} (\bibinfo{year}{2000}), \eprint{cond-mat/9906137}.

\bibitem[{\citenamefont{Feiguin et~al.}(2008)\citenamefont{Feiguin, Rezayi,
  Nayak, and Das~Sarma}}]{Feiguin08a}
\bibinfo{author}{\bibfnamefont{A.~E.} \bibnamefont{Feiguin}},
  \bibinfo{author}{\bibfnamefont{E.}~\bibnamefont{Rezayi}},
  \bibinfo{author}{\bibfnamefont{C.}~\bibnamefont{Nayak}}, \bibnamefont{and}
  \bibinfo{author}{\bibfnamefont{S.}~\bibnamefont{Das~Sarma}},
  \bibinfo{journal}{Phys. Rev. Lett.} \textbf{\bibinfo{volume}{100}},
  \bibinfo{pages}{166803} (\bibinfo{year}{2008}), \eprint{ar{X}iv:0706.4469}.

\bibitem[{\citenamefont{M\"{o}ller and Simon}(2008)}]{Moller08a}
\bibinfo{author}{\bibfnamefont{G.}~\bibnamefont{M\"{o}ller}} \bibnamefont{and}
  \bibinfo{author}{\bibfnamefont{S.~H.} \bibnamefont{Simon}},
  \bibinfo{journal}{Phys. Rev. B} \textbf{\bibinfo{volume}{77}},
  \bibinfo{pages}{075319} (\bibinfo{year}{2008}), \eprint{arXiv:0708.2680}.

\bibitem[{\citenamefont{Peterson
  et~al.}(2008{\natexlab{a}})\citenamefont{Peterson, Jolicoeur, and
  Das~Sarma}}]{Peterson08b}
\bibinfo{author}{\bibfnamefont{M.~R.} \bibnamefont{Peterson}},
  \bibinfo{author}{\bibfnamefont{T.}~\bibnamefont{Jolicoeur}},
  \bibnamefont{and}
  \bibinfo{author}{\bibfnamefont{S.}~\bibnamefont{Das~Sarma}},
  \bibinfo{journal}{Phys. Rev. Lett.} \textbf{\bibinfo{volume}{101}},
  \bibinfo{pages}{016807} (\bibinfo{year}{2008}{\natexlab{a}}),
  \eprint{ar{X}iv:0803.0737}.

\bibitem[{\citenamefont{Feiguin et~al.}(2009)\citenamefont{Feiguin, Rezayi,
  Yang, Nayak, and Das~Sarma}}]{Feiguin08b}
\bibinfo{author}{\bibfnamefont{A.~E.} \bibnamefont{Feiguin}},
  \bibinfo{author}{\bibfnamefont{E.}~\bibnamefont{Rezayi}},
  \bibinfo{author}{\bibfnamefont{K.}~\bibnamefont{Yang}},
  \bibinfo{author}{\bibfnamefont{C.}~\bibnamefont{Nayak}}, \bibnamefont{and}
  \bibinfo{author}{\bibfnamefont{S.}~\bibnamefont{Das~Sarma}},
  \bibinfo{journal}{Phys. Rev. B} \textbf{\bibinfo{volume}{79}},
  \bibinfo{pages}{115322} (\bibinfo{year}{2009}), \eprint{ar{X}iv:0804.4502}.

\bibitem[{\citenamefont{Peterson
  et~al.}(2008{\natexlab{b}})\citenamefont{Peterson, Park, and
  Das~Sarma}}]{Peterson08c}
\bibinfo{author}{\bibfnamefont{M.~R.} \bibnamefont{Peterson}},
  \bibinfo{author}{\bibfnamefont{K.}~\bibnamefont{Park}}, \bibnamefont{and}
  \bibinfo{author}{\bibfnamefont{S.}~\bibnamefont{Das~Sarma}},
  \bibinfo{journal}{Phys. Rev. Lett.} \textbf{\bibinfo{volume}{101}},
  \bibinfo{pages}{156803} (\bibinfo{year}{2008}{\natexlab{b}}),
  \eprint{ar{X}iv:0807.0638}.

\bibitem[{\citenamefont{Storni et~al.}(2008)\citenamefont{Storni, Morf, and
  Das~Sarma}}]{Storni08}
\bibinfo{author}{\bibfnamefont{R.~H.} \bibnamefont{Storni}},
  \bibinfo{author}{\bibfnamefont{R.~H.} \bibnamefont{Morf}}, \bibnamefont{and}
  \bibinfo{author}{\bibfnamefont{S.}~\bibnamefont{Das~Sarma}}
  (\bibinfo{year}{2008}), \eprint{ar{X}iv:0812.2691}.

\bibitem[{\citenamefont{d'Ambrumenil and Reynolds}(1988)}]{dAmbrumenil88}
\bibinfo{author}{\bibfnamefont{N.}~\bibnamefont{d'Ambrumenil}}
  \bibnamefont{and} \bibinfo{author}{\bibfnamefont{A.~M.}
  \bibnamefont{Reynolds}}, \bibinfo{journal}{J. Phys. C}
  \textbf{\bibinfo{volume}{21}}, \bibinfo{pages}{119} (\bibinfo{year}{1988}).

\bibitem[{\citenamefont{Peterson
  et~al.}(2008{\natexlab{c}})\citenamefont{Peterson, Jolicoeur, and
  Das~Sarma}}]{Peterson08a}
\bibinfo{author}{\bibfnamefont{M.~R.} \bibnamefont{Peterson}},
  \bibinfo{author}{\bibfnamefont{T.}~\bibnamefont{Jolicoeur}},
  \bibnamefont{and}
  \bibinfo{author}{\bibfnamefont{S.}~\bibnamefont{Das~Sarma}},
  \bibinfo{journal}{Phys. Rev. B} \textbf{\bibinfo{volume}{78}},
  \bibinfo{pages}{155308} (\bibinfo{year}{2008}{\natexlab{c}}),
  \eprint{ar{X}iv:0801.4819}.

\bibitem[{\citenamefont{Read and Rezayi}(1999)}]{Read99}
\bibinfo{author}{\bibfnamefont{N.}~\bibnamefont{Read}} \bibnamefont{and}
  \bibinfo{author}{\bibfnamefont{E.}~\bibnamefont{Rezayi}},
  \bibinfo{journal}{Phys. Rev. B} \textbf{\bibinfo{volume}{59}},
  \bibinfo{pages}{8084} (\bibinfo{year}{1999}), \eprint{cond-mat/9809384}.

\bibitem[{\citenamefont{Rezayi and Read}(2009)}]{Rezayi09a}
\bibinfo{author}{\bibfnamefont{E.~H.} \bibnamefont{Rezayi}} \bibnamefont{and}
  \bibinfo{author}{\bibfnamefont{N.}~\bibnamefont{Read}},
  \bibinfo{journal}{Phys. Rev. B} \textbf{\bibinfo{volume}{79}},
  \bibinfo{pages}{075306} (\bibinfo{year}{2009}), \eprint{cond-mat/0608346}.

\bibitem[{\citenamefont{Bonderson
  et~al.}(2009{\natexlab{c}})\citenamefont{Bonderson, Feiguin, M\"{o}ller, and
  Slingerland}}]{Bonderson09a}
\bibinfo{author}{\bibfnamefont{P.}~\bibnamefont{Bonderson}},
  \bibinfo{author}{\bibfnamefont{A.~E.} \bibnamefont{Feiguin}},
  \bibinfo{author}{\bibfnamefont{G.}~\bibnamefont{M\"{o}ller}},
  \bibnamefont{and} \bibinfo{author}{\bibfnamefont{J.~K.}
  \bibnamefont{Slingerland}} (\bibinfo{year}{2009}{\natexlab{c}}),
  \eprint{arXiv:0901.4965}.

\bibitem[{\citenamefont{Bonderson and Nayak}()}]{PB-CN-private}
\bibinfo{author}{\bibfnamefont{P.}~\bibnamefont{Bonderson}} \bibnamefont{and}
  \bibinfo{author}{\bibfnamefont{C.}~\bibnamefont{Nayak}},
  \bibinfo{note}{private communication}.

\bibitem[{\citenamefont{Xia et~al.}(2004)\citenamefont{Xia, Pan, Vicente,
  Adams, Sullivan, Stormer, Tsui, Pfeiffer, Baldwin, and West}}]{Xia04}
\bibinfo{author}{\bibfnamefont{J.~S.} \bibnamefont{Xia}},
  \bibinfo{author}{\bibfnamefont{W.}~\bibnamefont{Pan}},
  \bibinfo{author}{\bibfnamefont{C.~L.} \bibnamefont{Vicente}},
  \bibinfo{author}{\bibfnamefont{E.~D.} \bibnamefont{Adams}},
  \bibinfo{author}{\bibfnamefont{N.~S.} \bibnamefont{Sullivan}},
  \bibinfo{author}{\bibfnamefont{H.~L.} \bibnamefont{Stormer}},
  \bibinfo{author}{\bibfnamefont{D.~C.} \bibnamefont{Tsui}},
  \bibinfo{author}{\bibfnamefont{L.~N.} \bibnamefont{Pfeiffer}},
  \bibinfo{author}{\bibfnamefont{K.~W.} \bibnamefont{Baldwin}},
  \bibnamefont{and} \bibinfo{author}{\bibfnamefont{K.~W.} \bibnamefont{West}},
  \bibinfo{journal}{Phys. Rev. Lett.} \textbf{\bibinfo{volume}{93}},
  \bibinfo{pages}{176809} (\bibinfo{year}{2004}), \eprint{cond-mat/0406724}.

\bibitem[{\citenamefont{Pan et~al.}(2008)\citenamefont{Pan, Xia, Stormer, Tsui,
  Vicente, Adams, Sullivan, Pfeiffer, Baldwin, and West}}]{Pan08}
\bibinfo{author}{\bibfnamefont{W.}~\bibnamefont{Pan}},
  \bibinfo{author}{\bibfnamefont{J.~S.} \bibnamefont{Xia}},
  \bibinfo{author}{\bibfnamefont{H.~L.} \bibnamefont{Stormer}},
  \bibinfo{author}{\bibfnamefont{D.~C.} \bibnamefont{Tsui}},
  \bibinfo{author}{\bibfnamefont{C.}~\bibnamefont{Vicente}},
  \bibinfo{author}{\bibfnamefont{E.~D.} \bibnamefont{Adams}},
  \bibinfo{author}{\bibfnamefont{N.~S.} \bibnamefont{Sullivan}},
  \bibinfo{author}{\bibfnamefont{L.~N.} \bibnamefont{Pfeiffer}},
  \bibinfo{author}{\bibfnamefont{K.~W.} \bibnamefont{Baldwin}},
  \bibnamefont{and} \bibinfo{author}{\bibfnamefont{K.~W.} \bibnamefont{West}},
  \bibinfo{journal}{Phys. Rev. B} \textbf{\bibinfo{volume}{77}},
  \bibinfo{pages}{075307} (\bibinfo{year}{2008}), \eprint{ar{X}iv:0801.1318}.

\bibitem[{\citenamefont{Witten}(1989)}]{Witten89}
\bibinfo{author}{\bibfnamefont{E.}~\bibnamefont{Witten}},
  \bibinfo{journal}{Comm. Math. Phys.} \textbf{\bibinfo{volume}{121}},
  \bibinfo{pages}{351} (\bibinfo{year}{1989}).

\bibitem[{\citenamefont{Laughlin}(1983)}]{Laughlin83}
\bibinfo{author}{\bibfnamefont{R.~B.} \bibnamefont{Laughlin}},
  \bibinfo{journal}{Phys. Rev. Lett.} \textbf{\bibinfo{volume}{50}},
  \bibinfo{pages}{1395} (\bibinfo{year}{1983}).

\bibitem[{\citenamefont{Bonderson and Slingerland}(2008)}]{Bonderson07d}
\bibinfo{author}{\bibfnamefont{P.}~\bibnamefont{Bonderson}} \bibnamefont{and}
  \bibinfo{author}{\bibfnamefont{J.~K.} \bibnamefont{Slingerland}},
  \bibinfo{journal}{Phys. Rev. B} \textbf{\bibinfo{volume}{78}},
  \bibinfo{pages}{125323} (\bibinfo{year}{2008}), \eprint{arXiv:0711.3204}.

\bibitem[{\citenamefont{Haldane}(1983)}]{Haldane83}
\bibinfo{author}{\bibfnamefont{F.~D.~M.} \bibnamefont{Haldane}},
  \bibinfo{journal}{Phys. Rev. Lett.} \textbf{\bibinfo{volume}{51}},
  \bibinfo{pages}{605} (\bibinfo{year}{1983}).

\bibitem[{\citenamefont{Halperin}(1984)}]{Halperin84}
\bibinfo{author}{\bibfnamefont{B.~I.} \bibnamefont{Halperin}},
  \bibinfo{journal}{Phys. Rev. Lett.} \textbf{\bibinfo{volume}{52}},
  \bibinfo{pages}{1583} (\bibinfo{year}{1984}).

\bibitem[{\citenamefont{Milovanovic and Read}(1996)}]{Milovanovic96}
\bibinfo{author}{\bibfnamefont{M.}~\bibnamefont{Milovanovic}} \bibnamefont{and}
  \bibinfo{author}{\bibfnamefont{N.}~\bibnamefont{Read}},
  \bibinfo{journal}{Phys. Rev. B} \textbf{\bibinfo{volume}{53}},
  \bibinfo{pages}{13559} (\bibinfo{year}{1996}), \eprint{cond-mat/9602113}.

\end{thebibliography}

\end{document}